\documentstyle[prb,twocolumn,aps]{revtex}
\include{psfig}
\preprint{}
\begin{document}
\draft
\title{\bf
Nonequilibrium Josephson effect in mesoscopic ballistic multiterminal SNS
junctions}
\author{
P. Samuelsson, J. Lantz, V.S. Shumeiko and G.
Wendin}
\address{Department of Microelectronics and Nanoscience,
Chalmers University of Technology
and
G\"{o}teborg University, \\S-41296 G\"{o}teborg, Sweden}

%\date{\today}
\maketitle

\begin{abstract}

We present a detailed study of nonequilibrium Josephson currents and
conductance in ballistic multiterminal SNS-devices. Nonequilibrium is
created by means of quasiparticle injection from a normal reservoir
connected to the normal part of the junction. By applying a voltage at
the normal reservoir the Josephson current can be suppressed or the
direction of the current can be reversed. For a junction longer than
the thermal length, $L\gg\xi_T$, the nonequilibrium current increases
linearly with applied voltage, saturating at a value equal to the
equilibrium current of a short junction. The conductance exhibits a
finite bias anomaly around $eV \sim \hbar v_F/L$. For symmetric
injection, the conductance oscillates $2\pi$-periodically with the
phase difference $\phi$ between the superconductors, with position of
the minimum ($\phi=0$ or $\pi$) dependent on applied voltage and
temperature. For asymmetric injection, both the nonequilibrium
Josephson current and the conductance becomes $\pi$-periodic in phase
difference. Inclusion of barriers at the NS-interfaces gives rise to a
resonant behavior of the total Josephson current with respect to
junction length with a period $\sim \lambda_F$. Both three and four
terminal junctions are studied.

\end{abstract}

\section{Introduction}

The art of controlling Josephson current transport through mesoscopic
superconducting junctions poses many challenges for theory and
experiment from both fundamental and applied points of
view. \cite{BagwellSupM} Control of Josephson current requires
multi-terminal devices - superconducting transistors. One example is
the Josephson field effect transistor (JOFET), \cite{Houten,Akazaki} where
control of the Josephson {\em equilibrium} current is imposed via an
electrostatic gate. Another solution is to connect the normal region
to a normal voltage biased reservoir. Recent progress in fabrication
of superconducting junctions has brought forward a number of
interesting multiterminal structures, e.g. 2DEG,
junctions\cite{Akazaki,Takayanagi,Braginski} metallic junctions,
\cite{Morpurgo1,Baselmans} and high-Tc junctions. \cite{Lombardi}

Injection of electrons and holes allows nonequilibrium quasiparticle
distributions to be maintained in the N-region, making it possible to
control the {\em nonequilibrium} Josephson current. The problem of
nonequilibrium current injection in ballistic junctions is of
particular interest: the Josephson current is transported through
bands of Andreev levels.\cite{Andreev} This provides means for
achieving a dramatic variation of the Josephson current.

The purpose of this paper is to provide a broad description of
Josephson current transport through ballistic SNS junctions under
conditions of nonequilibrium in the normal region due to contact with
a voltage biased normal reservoir. Connection of the normal part of an
SNS junction to a normal electron reservoir gives rise to broadening
of the Andreev bound levels. Van Wees et al. \cite{Vanwees} were the
first ones to consider this broadening in perfect SNS junctions and
to describe essential aspects of the variation of the Josephson
current with voltage, in terms of nonequilibrium population of Andreev
levels. Moreover, Wendin and Shumeiko \cite{Wendin,Wendin2} have
predicted that nonequilibrium filling of Andreev levels may reveal
very large Josesphson currents with different directions, and that
pumping between levels could reverse the direction of the Josephson
current. This has been investigated in detail by Bagwell and coworkers
\cite{Bagwell} and by Samuelsson et al.
\cite{Samuelsson}.

Recently, Morpurgo et al. \cite{Morpurgo97b} observed Andreev levels
by using the injection lead as a spectroscopic probe. Suppression of
the Josephson current due to injection has been demonstrated in both
ballistic \cite{Braginski} and diffusive SNS junctions
\cite{Morpurgo1}. The physical mechanism of the effect in diffusive
junctions is essentially the same as in ballistic
junctions.\cite{Volkov95} Very recently, Baselmans
et. al. \cite{Baselmans} were able also to reverse the direction of
the Josephson current. 

A decisive step beyond the work of van Wees et al. \cite{Vanwees} was
taken by Samuelsson et al. \cite{Samuelsson}, who showed that an
essential aspect is the ability of the scatterer at the injection
point to shift the phases of the quasiparticles. In such a case, the
connection to the injection lead also affects the form of the wave
function of the Andreev resonances, and therefore affects Josephson
currents flowing through the resonances. As a result, modification of
the Josephson current under injection does not reduce to the effect of
non-equilibrium population. This is particularly dramatic for long
junctions, where the equilibrium Josephson current is exponentially
small at finite temperature. \cite{Kulik}

In contrast, this {\em anomalous} nonequilibrium Josephson current
does not depend on the length of the junction (long-range Josephson
effect). This means that, in principle, a dissipationless current of
the order of the equilibrium Josephson current of a short junction can
be restored under conditions of filling up all the Andreev levels in
the gap. The effect is most pronounced in junctions with a small
number of transport modes. This opens up the possibility for a new
kind of Josephson transistor where the supercurrent is turned on when
the gate voltage is switched from $eV=0$ to $eV=\Delta$.\cite{Patent}

The complete picture of the nonequilibrium current also includes the
current injected into the junction. This {\em injection} current is
dependent on the properties of the Andreev levels in the junction. It
therefore provides information on the nonequilibrium Josephson
current. We found that it is closely related to the anomalous current
and has similar properties. The injection current has also in itself
been at the focus of great interest in recent literature.
\cite{Lambertrew}

The paper is organized as follows. In section $II$ we present a
general discussion of the currents in a 3-terminal SNS device. In
section $III$ we describe our model based on the stationary BdG
equation. We derive all currents in the case of a three terminal
junction without barriers at the NS-interface in section $IV$. In
section $V$ we discuss the equilibrium and nonequilibrium Josephson
currents, both in a short and long junctions. The effect of barriers
at the NS-interfaces is discussed in section $VI$ and the injection
current and the conductance are analyzed in section $VII$. In section
$VIII$ we discuss the four terminal junction and how it differs from
the three terminal one. Finally, in section $IX$ we present our
conclusions.

\section{Nonequilibrium Josephson currents}

\begin{figure}
\centerline{\psfig{figure=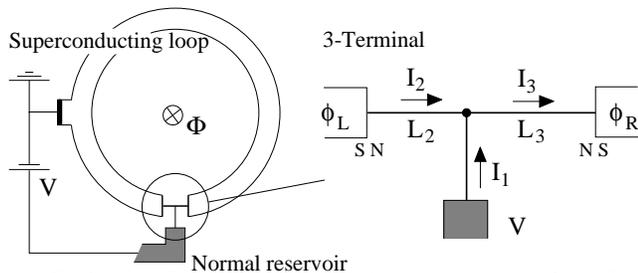,height=3.5cm}}
\caption{A schematic picture of the three terminal SNS junction setup
under consideration, with a normal reservoir attached to the normal
part of the junction. The normal reservoir is connected to the
superconducting loop (grounded) via a voltage source biased at
$V$. The right figure shows a close-up of the junction area with the
arrows showing the direction of the current flow in the junction}
\label{juncfig}
\end{figure}

We will consider two junction configurations: 3- and 4-terminal (see
Fig. \ref{juncfig}). The normal part of the junction is inserted between
two superconducting electrodes. The superconducting electrodes are
connected with each other to form a loop and the magnetic flux threading the
loop allows us to control the phase difference $\phi=\phi_R-\phi_L$ across
the junction.

We consider a junction in the ballistic limit, i.e when the length
$L=L_2+L_3$ of the normal part of the junction is shorter then both
the elastic and inelastic scattering lengths, $L\ll l_e,l_i$. The
3-terminal configuration is an elementary structure which gives all
necessary information for understanding also the properties of the
4-terminal junction, to be discussed below. We use a simplified
description of the connection point, modeling it by a scattering
matrix $S$ that connects ingoing and outgoing wave function
amplitudes \cite{Buttiker}
\begin{equation}
\Psi_{out} = S \Psi_{in},
\end{equation}
with

\begin{equation}
S=\left( \begin{array}{lll}
\sqrt{1-2\epsilon} & \sqrt\epsilon& \sqrt\epsilon\\
         \sqrt\epsilon & r & d\\
                  \sqrt\epsilon & d & r\\
\end{array} \right),
\label{scatmat}
\end{equation}
where $r$ and $d$ are reflection and transmission amplitudes for
scattering between lead 2 to lead 3 and $\sqrt{\epsilon}$ is the
scattering amplitude from the injection lead 1 to lead 2 or 3. In a
multichannel treatment, $r,d$ and $\epsilon$ become matrices
describing the scattering between the channels. In this paper we
however choose to consider a single-mode structure.

In the junction presented in Fig. \ref{juncfig}, the current $I_1$
injected into the junction from the normal reservoir splits at the
connection point. At the NS-interfaces, the normal current is
converted into a supercurrent. The supercurrent flows around the loop
and is drained at a point connected to the normal reservoir via a
voltage source biased at voltage $V$.  There are two major questions
about the currents: (i) what is the current $I_1$ in injection
electrode 1 as function of the applied voltage, and (ii) how is the
current split between the arms 2 and 3. The first problem has been
discussed earlier, \cite{Nakano,Kadigrobov} the picture is the
following: due to Andreev quantization the problem is equivalent to a
resonant transmission problem. For weak coupling to the normal
reservoir, $\epsilon \ll 1$, the probability of an incoming electron
to be reflected is large unless its energy coincides with an Andreev
level. In such a case, the electron is back scattered as a hole which
produces a current density peak. The current as a function of applied
voltage between the normal reservoir and the junction (IVC) thus
increases stepwise, typical for resonant transport, with position and
height of the steps depending on the phase difference between the
superconductors.

The current distribution among the left and right arms of the junction
is also phase dependent. However, there is a less trivial aspect of
the problem related to the Josephson current in the loop. There is no
possibility to distinguish the Josephson current which flows along the
loop (as the result of an applied phase difference) from the split
injection current {\em except} in the limit of weak coupling to the
external reservoir. In the limit $\epsilon \ll 1$ the injection
current $(\sim\epsilon)$ vanishes while the Josephson current remains
finite. This allows us to separate the problem of the Josephson
current under injection from the problem of splitting of the injection
current.

\begin{figure}
\centerline{\psfig{figure=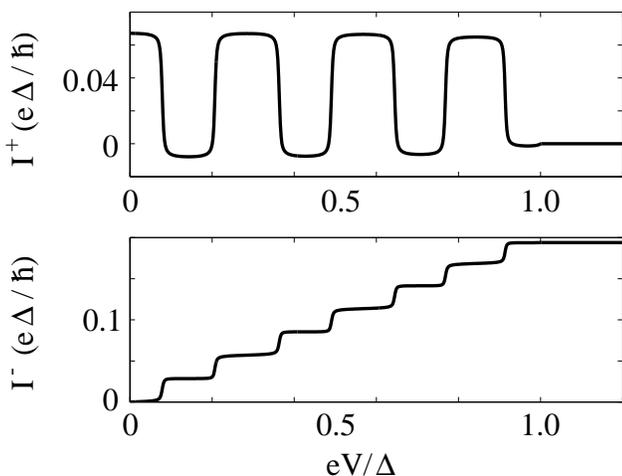,height=6.5cm}}
\caption{The current voltage characteristics (IVC) for $I^+$ (upper) and
$I^-$ (lower) for a junction with seven Andreev levels for $0<E<\Delta$.
The currents jump every time the voltage $eV$ is equal to the energy of an
Andreev level, typical for resonant transport}
\label{spectro}
\end{figure}

The scattering states carrying the current can qualitatively be
described as electrons or holes entering the SNS junction from the
injection lead 1, being split at the connection point, scattered back
and forth in the junction by Andreev reflections at the NS-interfaces
and normal reflections at the connection point, and then finally
leaving the junction, having effectively transported current from one
superconductor to the other. When the lifetime of the Andreev
resonances is smaller than the inelastic scattering time in the
junction, the quasiparticle distribution in the normal region is
determined by the Fermi distribution function of the normal reservoir,
and the current in the leads $j=2$ or $3$ from injected quasiparticles
can be written

\begin{equation}
I_j=\int_{-\infty}^{\infty} dE (i^e_j n^e +i^h_jn^h)
\end{equation}
where $i_j^{e(h)}$ is the current density for injected electrons
(holes) and $n^{e(h)}=n_F(E \pm eV)$ are the Fermi distribution
functions in the normal reservoir, with
$n_F=[1+\mbox{exp}(E/kT)]^{-1}$. This current can  conveniently be rewritten

\begin{equation}
I_j=\int_{-\infty}^{\infty}
dE\left[\frac{i^+_j}{2}(n^e+n^h)+\frac{i^-_j}{2}(n^e-n^h)\right]=I^+_j+I^-_j
\label{curreq}
\end{equation}
where $i^+=i^e+i^h$ and $i^-=i^e-i^h$. Quasiparticles are also
injected from the superconductors for energies above the
superconducting gap. Since the superconductors are grounded ($V=0$),
the current from the superconductors is an equilibrium current. This
current plus the current $I^+=I_2^+= I_3^+$ injected from the normal
reservoir in absence of applied voltage, is the total {\em equilibrium}
current. Applying a bias voltage $(V\ne0)$, $I^+$ becomes the {\em
nonequilibrium} current due to population of the empty Andreev levels,
giving rise to current jumps when the injection energy $eV$ equals the
Andreev level energies (see Fig. \ref{spectro}). This makes it
possible to probe the energy of the Andreev levels.
\cite{Vanwees,Samuelsson,Bagwell}

The $I^-$ part of the current is entirely nonequilibrium current. It partly
consists of the injection current; however, there is also a component which
does not vanish in the limit of weak coupling to the reservoir: we call
this the {\em anomalous Josephson current}. \cite{Samuelsson} This current
results from a different form of the Andreev resonance wave functions in
the {\em open} junction compared with the wave functions of true Andreev
{\em bound} states.
The origin of the anomalous current can qualitatively be described by
considering the lowest order quasiparticle classical paths which contribute
to the resonances in transparent junctions ($R\ll1$) with perfect NS
interfaces.

\begin{figure}
\centerline{\psfig{figure=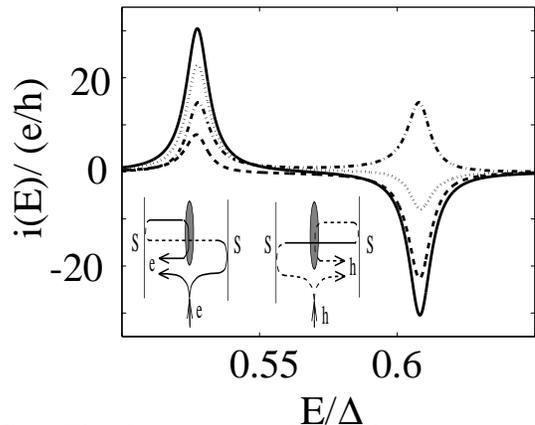,height=6.5cm}}
\caption{The charge current density for two resonant Andreev levels
for injected electrons $i^e$ (dotted) and holes $i^h$ (dashed), their
sum $i^+$ (solid) and difference $i^{-}$ (dash-dotted). Note that the
difference current $i^-$ has the same sign for both resonances. Inset:
Two lowest order paths for an injected electron (solid) or a hole
(dashed) at a resonance. The grey ellipse denotes the effective
scatterer due to the three lead connection. The difference of the
currents due to these processes is proportional to
$\mbox{Im}(rd^*)\sin\phi$, the first order term of the anomalous
current }
\label{ccd2}
\end{figure}

Consider a resonant state where the most of the electrons move to the
left and the holes to the right, only a fraction of them travellling
in the opposite direction due to normal scattering at the connection
point. An injected electron gives rise to a leftgoing electron in lead
2 with the amplitude $1+e^{i\phi_R}d^*e^{-i\phi_L}r$ with
$\phi=\phi_R-\phi_L$ (not taking electron and hole dephasing and the
energy dependent phase picked up when Andreev reflecting into account)
thus giving a contribution to the current of the order
$1+RD+\mbox{Re}(rd^*e^{i\phi})$ (see inset in
Fig. \ref{ccd2}). Correspondingly, an injected hole gives rise to a
rightgoing hole in lead 3 with amplitude $1+e^{-i\phi_L} de^{i\phi_R}
r^*$ and a contribution to the current of order
$1+RD+\mbox{Re}(rd^*e^{-i\phi})$ (see right figure in inset in
Fig. \ref{ccd2}). The difference current $i^-$ thus contains a part
proportional to
$\mbox{Re}[rd^*(e^{i\phi}-e^{-i\phi})]=2\mbox{Im}(rd^*)\sin(\phi)$,
which is the leading term in the anomalous current. At a resonant
state where the particles move in the opposite direction, i.e the
electrons to the right and the holes to the left, we find from the
same arguments that the anomalous current is again proportional to
$2\mbox{Im}(rd^*)\sin(\phi)$, {\em with the same sign}. The anomalous
current thus flows in the same direction for all resonances, in
contrast to the equilibrium Josephsson current which changes sign from
one level to the next. The IVC for $I^-$ is thus a staircase, as shown
in Fig. \ref{spectro}, saturating at $eV>\Delta$ due to the absence of
sharp resonances for energies above the superconducting gap. This has
a dramatic effect on the long range properties of the Josephson
current.

For a long junction $(L\gg \xi_0=\hbar v_F/\Delta)$, the IVC in Fig.
\ref{spectro} becomes dense, since there is a large number $\sim
L/\xi_0$ of Andreev levels in the junction. The spacing between the
Andreev levels is $\sim \hbar v_F/L$, so at temperatures exceeding
the interlevel distance, the current $I^+$ is averaged to zero while
$I^-$ is reduced to a smooth ramp function. We thus get a current
$I^-$ that increases linearly with voltage up to $eV=\Delta$ and
saturates at a level of the order of the equilibrium Josephson current
of a short junction, $I\sim e\Delta/\hbar$. This current is
independent of the length of the junction, since there is a large
number of levels $\sim L$ each carrying a current $\sim 1/L$.

\section{Calculation of the current}
\subsection{General formulation}

We consider a three-terminal junction with asymmetric current
injection ($L_2\neq L_3$) and perfect transmission at the NS
interfaces. The junction can be described by the stationary 1-D
Bogoliubov-de Gennes (BdG) equation\cite{Degennes}

\begin{equation}
\label{bdgeq}
\left[ \begin{array}{cc}
        H_{0}&\Delta\\
        \Delta^*&-H_{0}
\end{array} \right] \Psi=E \Psi \hspace{1cm}
H_{0}=-\frac{\hbar^{2}}{2m}\frac{d^2}{dx^2}-E_{F}
\end{equation}
which gives $E$ as a departure from $E_{F}$. We apply the approximation
\cite{Likharev} with $\Delta (x)$ constant in the
superconductors and zero in the normal region.

\begin{equation}
\Delta(x)= \left\{ \begin{array}{ll}
        \Delta e^{i\phi_L} & x<-L_2 \\
        0 & -L_2 <x<L_3 \\
        \Delta e^{i\phi_R} & x>L_3
        \end{array}
        \right. ,
\end{equation}
where the phase difference between the superconductors is
$\phi=\phi_R-\phi_L$. We can then make an ansatz with plane waves in
the different regions of the junction. For positive energies $E>0$ we
put in the normal regions $j=1,2,3$,

\begin{eqnarray}
\Psi_{j}&=&c^{+,e}_{j}\left[\begin{array}{c} 1\\ 0 \end{array}\right] e^{i
k^ex}+c^{h,-}_{j}\left[\begin{array}{c} 0 \\ 1 \end{array}\right]
e^{-ik^hx} \nonumber \\
&& +c^{e,-}_{j}\left[\begin{array}{c} 1\\ 0 \end{array}\right]
e^{-ik^ex}+c^{h,+}_{j}\left[\begin{array}{c} 0\\ 1
\end{array}\right] e^{ik^hx}
\end{eqnarray}
and in the superconductors $j=L,R$

\begin{eqnarray}
\Psi_{j}&=&d^{e,+}_{j}\left[\begin{array}{c} ue^{i\phi_j}\\ v
\end{array}\right] e^{i q^ex}+d^{h,-}_{j}\left[\begin{array}{c}
ve^{i\phi_j}\\ u \end{array}\right] e^{-iq^hx} \nonumber \\
&& +d^{e,-}_{j}\left[ \begin{array}{c} ue^{i\phi_j}\\ v
\end{array}\right] e^{-iq^ex}+d^{h,+}_{j}\left[\begin{array}{c}
ve^{i\phi_j}\\ u \end{array}\right] e^{iq^ex}.
\end{eqnarray}
The coherence factors $u$ and $v$ are defined as

\begin{equation}
u(+),v(-)= \left\{ \begin{array}{ll}
        \sqrt{\frac{1}{2}(1\pm\xi/E)} & E>\Delta \\
        \sqrt{\frac{1}{2}(E\pm\xi)/\Delta} & E<\Delta
        \end{array}
        \right.
\end{equation}
where $\xi=\sqrt{E^2-\Delta^2}$ for $E>\Delta$ and
$\xi=i\sqrt{\Delta^2-E^2}$ for $E<\Delta$. The wavevectors are
$q^{e,h}=\sqrt{2m/\hbar^2}\sqrt{E_{F}\pm \xi}$ in the superconductors
and and $k^{e,h}=\sqrt{2m/\hbar^2}\sqrt{E_{F}\pm E}$ in the normal
regions. The wavefunctions are matched at the NS-interfaces and at the
injection point. The three-terminal injection point is modeled by the
scattering matrix\cite{Buttiker,Nakano} given by
Eq. (\ref{scatmat}). The scattering amplitudes $\epsilon$ ($0\leq
\epsilon\leq 0.5$), $d$ and $r$ obey the relations
$Re(rd^\ast)=-\epsilon/2$ and $D+R=1-\epsilon$ ($D=|d|^2, R=|r|^2$)
due to the unitarity of the scattering matrix.  Moreover,
$\mbox{Im}(rd^*)=\sigma\sqrt{RD-\epsilon^2/4}$, with $\sigma=\pm 1$
dependent on the phase of the scatterer. For simplicity the coupling
parameter $\epsilon$ is chosen real and positive.  The scattering
amplitudes are assumed to be energy independent, which gives the
scattering matrix for hole wavefunction amplitudes $S_h=S_e^*$.

Assuming $\Delta \ll E_F$ we make the approximation $q^{e}=q^{h}=k_{F}$
in the superconductors and $k^{e}=k^{h}=k_{F}$ in the normal region
except in exponentials where we put $k^{e,h}=k_{F}\pm E/(\hbar
v_F)$. At energies $E<\Delta$, only electrons and holes from the normal
reservoirs are injected in the junction. For $E>\Delta$ quasiparticles
from the superconductors are also injected. The current density in the
three normal regions, which is what is needed to calculate all
currents in the junction, are calculated using the quantum mechanical
formula \cite{BTK}

\begin{equation}
i_j(E)=\frac{e}{h}(|c_j^{+,e}|^2-|c_j^{-,e}|^2-|c_j^{+,h}|^2+|c_j^{-,h}|^2).
\label{btkcurr}
\end{equation}

We now define energy dependent phases
$\theta_{2,3}=\gamma-\beta_{2,3}$ in each of the leads $2$ and $3$,
consisting of the phase $\gamma=\arccos(E/\Delta)$ picked up by the
electrons and holes when Andreev reflecting, and the dephasing
$\beta_{2,3}=(k^e-k^h)L_{2,3}=2EL_{2,3}/(\hbar v_F)$ of the electrons
and holes while propagating ballistically through the normal
region. Furthermore, it is convenient to separate out the specific
features of asymmetry by introducing sum phases
2$\theta=\theta_2+\theta_3$, $\beta = \beta_2+\beta_3$, and the
difference phases $\chi=\theta_2-\theta_3$, defining essential phase
parameters characterizing the junction,

\begin{eqnarray}
\theta = \gamma-\beta/2 = \arccos(E/\Delta) - EL/(\hbar v_F) \\
\label{thetadef}
\chi = \beta_3-\beta_2 = 2El/(\hbar v_F)
\label{chidef}
\end{eqnarray}
where $L=L_2+L_3$ and $l= L_3-L_2$

The current densities of the scattering states in leads $2$ and $3$ from
electrons $i_{2,3}^e$ and holes $i_{2,3}^h$ are then given by

\begin{eqnarray}
i_2^{e,h}&=&-{e\over h}{\epsilon\over Z} \left\{2D\sin\phi\sin2\theta
\right.  \nonumber \\ && \left. \pm \left[\sigma
2\sqrt{RD-\epsilon^2/4}\sin\phi(\cos\chi-\cos2\theta)
\right. \right. \nonumber \\ && \left. \left. +\epsilon
\left[1-\cos(2\gamma-\beta_2) + \cos\phi (\cos\chi-\cos2\theta)
\right]\right]\right\}
\label{ieh2}
\end{eqnarray}

\begin{eqnarray}
i_3^{e,h}=&&-{e\over h}{\epsilon\over Z}\left\{2D\sin\phi\sin2\theta
\right.  \nonumber \\ && \left. \pm
\left[2\sigma\sqrt{RD-\epsilon^2/4}\sin\phi(\cos\chi-\cos2\theta)
\right. \right.  \nonumber \\ && \left. \left. -\epsilon
\left[1-\cos(2\gamma-\beta_3) + \cos\phi (\cos\chi-\cos2\theta)
\right]\right]\right\}
\label{ieh3}
\end{eqnarray}
where

\begin{equation}
Z=[(1-\epsilon)\cos2\theta -R\cos\chi -D\cos\phi]^2+\epsilon^2\sin^22\theta
\label{Z}
\end{equation}
From Eqs. (\ref{ieh2}) and (\ref{ieh3}) it follows that the sum of the
electron and hole current densities, $i^+=i^e+i^h$, are equal in leads 2 and 3,
giving the sum current density

\begin{equation}
i^+=i_3^+=i_2^+=-{4e\over h}{\epsilon\over
Z}\left\{D\sin\phi\sin2\theta\right\}.
\label{sumcurr}
\end{equation}
The difference current densities $i^-=i^e-i^h$ in leads 2 and 3 are
not equal, however. We therefore define the anomalous current density
$i_a$ as that part of the difference current density which survives in
the limit $\epsilon\rightarrow 0$,

\begin{equation}
i_a=-\sigma \frac{4e}{h}\frac{\epsilon}{Z}\left\{\sqrt{RD-\epsilon^2/4}
\sin\phi(\cos\chi-\cos2\theta) \right\}
\label{avdiffcurr}
\end{equation}
The injection current density $i_{inj}=i_3^- -i_2^-$ is given by,

\begin{equation}
i_{inj}=\frac{4e}{h}\frac{\epsilon^2}{Z}\left\{\sin^2\chi + (\cos\chi
+\cos\phi) (\cos\chi-\cos2\theta) \right\}
\label{injdiffcurr}
\end{equation}
and splits asymmetrically between the two horizontal arms $2$ and $3$,

\begin{equation}
i_{inj2,3}=\pm \frac{2e}{h}\frac{\epsilon^2}{Z}\left\{1-\cos(2\theta-\beta_{2,3}).
+ \cos\phi(\cos\chi-\cos2\theta) \right\}
\end{equation}
From the relations $i^{+}(E)=-i^{+}(-E)$ and $i^{-}(E)=i^{-}(-E)$ one
can calculate the current densities for all energies inside the gap
$|E|<\Delta$. The continuum current density, for energies outside the
gap $|E|>\Delta$, is calculated in the same way. However, since the
Andreev reflection probability decays very rapidly outside the gap,
the Andreev resonances become very broad and contribute much less to
the current. Only the quasiparticles injected from the superconductors
contribute significantly to the current, as will be discussed
below. The full formulas for the continuum current density for a
symmetric junction $l=0$ is presented in Appendix A.

\subsection{Weak coupling limit}

\begin{figure}
\centerline{\psfig{figure=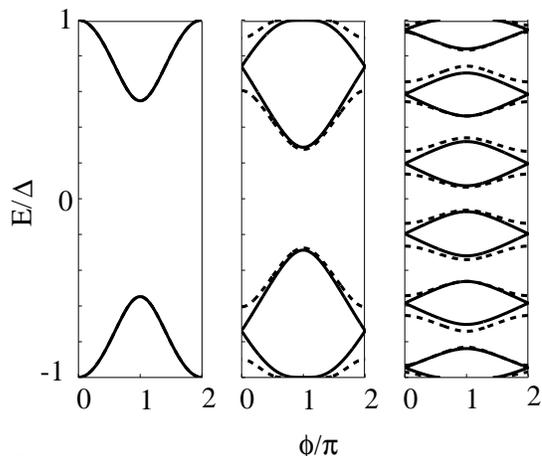,height=6.5cm}}
\caption{Andreev bound state energies as a function of phase difference
$\phi$ for different lengths $L=0$ (left), $L\sim \xi_0$ (middle) and
$L\gg\xi_0$ (right) of the junction with $D=0.7$. Solid lines are for a symmetric
junction $l=0$, dashed for an asymmetric one. A gap opens up in the
spectrum at $\phi=0$ due to the asymmetry.}
\label{asandr}
\end{figure}

Throughout the paper we will mainly discuss the situation when the normal reservoir is weakly coupled to the normal part of the junction, $\epsilon \ll 1$. In this limit the Andreev resonances are very sharp and the current densities are calculated by evaluating the expression $\epsilon/Z$ appearing in the Eqs. (\ref{sumcurr})-(\ref{injdiffcurr}), in the limit $\epsilon \rightarrow 0$. This is done in detail in Appendix B, and gives [see Eq. (\ref{epszeta})]

\begin{equation}
\lim_{\epsilon\rightarrow 0} \frac{\epsilon}{Z}=\sum_{n,\pm} \frac{\pi}{D|\sin \phi \sin 2\theta|}\left|\frac{dE}{d\phi}\right| \delta(E-E_n^{\pm}).
\label{epszero}
\end{equation}
where $E_n^{\pm}$ are the energies of the bound Andreev states. To
calculate the current density, information about the bound state
energies as well as the derivative of the energy with respect to phase
difference is thus needed. The bound state energies are given by the
zeros of the denominator $Z$ [Eq. (\ref{Z})] at $\epsilon=0$, namely
\cite{Bagwell4}

\begin{equation}
\cos2\theta = R\cos\chi + D\cos\phi.
\label{bsteq}
\end{equation}
The energy of the Andreev levels as a function of phase difference
$\phi$ is plotted in Fig. \ref{asandr}. In the figure it is shown that
the Andreev levels appear in pairs, labeled by $n$, with an upper
($+$) and a lower ($-$) level (referring to $E>0$). The index $n$ is
zero for the pair of levels with positive energy closest to $E_F$. In
the case of one single bound state, the level is labeled by $E_0^-$.

The derivative of the bound state energy with respect to phase is
obtained by differentiating Eq. (\ref{bsteq}), giving

\begin{eqnarray}
&&\frac{dE_n^{\pm}}{d\phi}=
\frac{D\sin\phi}{2\sin2\theta} \nonumber \\ 
&& \times \left( \frac{1}{\sqrt{\Delta^2-(E_n^{\pm})^2}}
+\frac{L}{\hbar v_F } + \frac{l}{\hbar v_F} R
\frac{\sin\chi}{\sin2\theta}\right)^{-1}.  
\label{asymdedfi}
\end{eqnarray}
The expression for the sum current density is given by inserting Eqs. (\ref{epszero})-(\ref{asymdedfi}) into Eq. (\ref{sumcurr}), giving

\begin{equation}
i^+=\frac{2e}{\hbar}\sum_{n,\pm}\frac{dE}{d\phi}\delta(E-E_n^{\pm}),
\label{sumepsgotozero}
\end{equation}
where the relation $\mbox{sgn}[(dE/d\phi) \sin\phi \sin 2\theta]=-1$
[see Eq. (\ref{signrel})] has been taken into account. The expression
(\ref{sumepsgotozero}) coincides with the equation for the Andreev
bound state current \cite{Vitaly1} derived directly from the BdG
equation.  From the alternating slopes of the energy-phase relation
$E(\phi)$, plotted in Fig. \ref{asandr}, it is clear that the sum
current density ($\sim dE/d\phi$) changes sign between two subsequent
Andreev resonances (see Fig. \ref{ccd2}).

The anomalous current density $i_a$ is given directly by inserting Eq. (\ref{epszero}) into (\ref{avdiffcurr}), namely

\begin{eqnarray}
i_a&=&-\sigma {2e\over\hbar} \mbox{sgn}(\sin\phi)\sqrt{RD} \nonumber \\
&& \times \sum_{n,\pm} \frac{\cos\chi -\cos\phi}{|\sin2\theta|} \left|{dE_n^{\pm} \over d\phi}\right| \delta(E-E_n^{\pm}).
\label{ianzero}
\end{eqnarray}
For a symmetric junction $l=0, \cos\chi=1$, the anomalous current
density does not change sign as a function of energy, opposite to the
sum current density (see Fig. \ref{ccd2}). For finite asymmetry, the
anomalous current might change sign. However, this does not lead to
strong suppression of the total anomalous current, as will be shown
below in section VB.

The injection current $i_{inj}=i_3^- -i_2^-$ is proportional to
$\epsilon ^2$ and thus goes to zero for $\epsilon \ll 1$. We
approximate the injection current in the weak coupling limit by the
first order term in $\epsilon$, given by inserting the expression for
$\epsilon/Z$ in the zero coupling limit into Eq. (\ref{epszero})

\begin{eqnarray}
i_{inj}&&= \epsilon{8e\over \hbar} \sum_{n,\pm}
\frac{\sin^2\chi + D(\cos\chi -\cos\phi)^2}{|\sin2\theta|}
\nonumber \\
&& \times \left |\frac{dE_n^{\pm}}{d\phi}\right| \delta(E-E_n^{\pm}).
\label{injcurrdens}
\end{eqnarray}
The injection current density is closely related to the anomalous
current density $i_a$, in the sense that the injection current density
is positive for all energies and values of the phase difference
$\phi$.

\subsection{Structure of the nonequilibrium current}

Including the continuum contribution from the superconductors
(Appendix A) in Eq.  (\ref{curreq}), we can finally write down the
structure of the total current in each lead:

\begin{eqnarray}
I_j = \int_{-\infty}^{\infty}dE
\left[\frac{i_j^+}{2}(n^e+n^h)+\frac{i_j^-}{2}(n^e-n^h)+i^s n_F \right],
\label{totcurr}
\end{eqnarray}
where $i^s$ is the current density from the quasiparticles injected
from the superconductors. The equilibrium current $(V=0)$ flowing in
leads $2$ and $3$ is given by

\begin{equation}
I_{eq}=\int dE \left[ i^+ + i^s \right] n_F
\label{ieqdef}
\end{equation}
while in lead $1$ it is zero. Subtracting the equilibrium current from
the total current we get the {\em nonequilibrium} current in the horizontal
leads $2$ and $3$. We divide the nonequilibrium current into the
the {\em regular current} $I_r$ associated with the nonequilibrium
population of the existing resonant states,

\begin{equation}
I_r=\int dE \left[\frac{i^+}{2}(n^e+n^h-2n_F) \right],
\label{irdef}
\end{equation}
the {\em anomalous current} $I_a$ associated with the essential modification
of the Andreev states due to the open normal lead,

\begin{equation}
I_a=\int dE \left[\frac{i_a}{2}(n^e-n^h) \right],
\label{iadef}
\end{equation}
and the injected current $I_1$

\begin{equation}
I_1= I_{inj}=\int dE \left[\frac{i_{inj}}{2}(n^e-n^h) \right].
\label{i1def}
\end{equation}
With these definitions, the total currents in leads 2 and 3 may be
written as

\begin{eqnarray}
I_2=I_{eq}+I_r+I_a-I_{inj,2}, \\
I_3=I_{eq}+I_r+I_a+I_{inj,3}, \nonumber
\end{eqnarray}
where $I_{inj}= I_{inj,2}+I_{inj,3}$. As discussed in Section II, the
separation of the anomalous current is arbitrary, and has physical
meaning only in the weak coupling limit when $I_{inj} \rightarrow 0$.

In the weak coupling limit, the integrals in Eqs.
(\ref{irdef})-(\ref{iadef}) become sums over resonant states 
\begin{equation}
I_r=\frac{e}{\hbar}\sum_{n,\pm}\frac{dE_n^{\pm}}{d\phi}\left[n^e(E_n^{\pm})+
n^h(E_n^{\pm})-2n_F(E_n^{\pm})\right],
\label{irbound}
\end{equation}

\begin{eqnarray}
I_a&=&-\sigma {e\over\hbar} \mbox{sgn}(\sin\phi)\sqrt{RD}  \nonumber \\
&& \times \sum_{n,\pm} \frac{\cos\chi -\cos\phi}{|\sin2\theta|} \left|\frac{dE_n^{\pm}}{d\phi}\right|\left[n^e(E_n^{\pm})-n^h(E_n^{\pm})\right].
\label{iabound}
\end{eqnarray}
The equilibrium current for energies $|E|<\Delta$ is given by inserting Eq. (\ref{sumepsgotozero}) into (\ref{ieqdef}), 

\begin{equation}
I_{eq}^{b}=\frac{2e}{\hbar}\sum_{n,\pm}\frac{dE_n^{\pm}}{d\phi}n_F(E_n^{\pm}),
\label{sumboundstate}
\end{equation}
For energies above the gap, the equilibrium current results from quasiparticles injected from the superconductors only, since this current is the only continuum current being finite in the weak coupling limit (see Appendix A). 

\section{Josephson current of a short junction}

For a short junction $L=l=0$, there is exactly one resonance for
positive energies $0<E<\Delta$. For no coupling to the normal
reservoir $\epsilon=0$, this resonant Andreev state is converted into
a bound Andreev state, with the dispersion relation $E_0^-=\Delta
\sqrt{1-D\sin^2(\phi/2)}$. The equilibrium current of a short junction
is thus given by the well known
\cite{Furusaki} relation

\begin{equation}
I_{eq}=\frac{e\Delta}{\hbar}\frac{D \sin \phi}{2\sqrt{1-D\sin^2(\phi/2)}}
\tanh(E_0^-/2kT).
\label{shortieq}
\end{equation}
The continuum current is zero, which can be seen by putting $L=0$
$(\beta=0)$ in the equations for the continuum current in Appendix
A. At zero temperature and zero applied bias, only the level with
negative energy $-E_0^-$ is populated. For an applied a voltage bias
$V>0$, the electron (hole) population is shifted upwards (downwards)
in energy. When the voltage $eV=E_0^-$, the energy of the resonant
level, the level becomes populated and there is an abrupt jump of the
current. The regular part of the current, $I_r$, jumps an amount
$\delta I_r=-I_{eq}$, thus cancelling the equilibrium Josephson
current. This has recently been observed in experiments.
\cite{Braginski,Morpurgo1}

The anomalous current jumps by the amount

\begin{figure}
\centerline{\psfig{figure=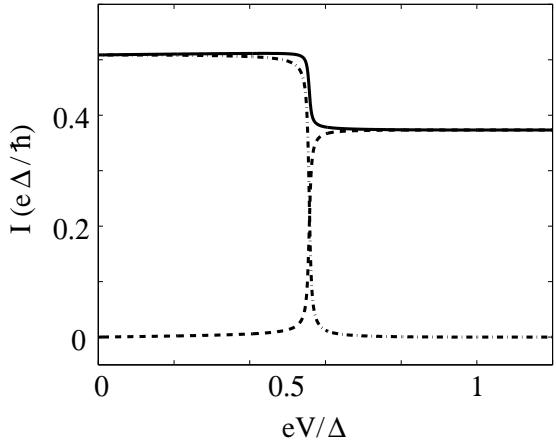,height=6.5cm}}
\caption{The currents $I_{eq}+I_r$ (dash-dotted), $I_a$ (dashed) and the
total Josephson current $I_{eq}+I_r+I_a$ (solid) in the horizontal leads
$2$ and $3$ as a function of voltage $V$ at $T=0$ for a short
junction $L=0$ with $D=0.8, \phi=3\pi/4, \epsilon=0.01$ and $\sigma=-1$.
The total current is $I_{eq}$ for $eV<E_0^-$ and $\delta I_a$ for
$eV>E_0^-$.}
\label{shortIVC}
\end{figure}
The effect of finite temperature in a zero length junction is merely to
smear the steps in the IVC.

In the symmetric case ($l=0)$ it is interesting to extend the
discussion to a longer junction with two resonant levels (see
Fig. \ref{asandr}), since the current distribution between the levels
becomes nontrivial. \cite{Wendin,Bagwell} In the limit $D\ll1$, both
resonances have energies close to the gap edge, $E_0^{\pm} \approx
\Delta$, and with the additional approximation $\beta/2>\sqrt{D}$ we
obtain the expression for the derivative of energy with respect to
phase [see Eq. (\ref{asymdedfi})]

\begin{equation}
\frac{dE_0^{\pm}}{d\phi}=\pm\frac{\Delta\sqrt{D}}{4}\frac{L}{\xi_0}\frac{\sin(\phi)}{|\sin(\phi/2)|\sqrt{1-D\sin^2(\phi/2)}}.
\end{equation}
The equilibrium bound state current becomes proportional to $I_{eq}^b \sim
dE_0^+/d\phi+dE_0^-/d\phi\sim D$ (taking terms of order $D$ into account),
but for the currents of the individual levels $\sim \sqrt{D}$. The resonant
levels thus carry opposite ``giant'' currents which almost cancel in
equilibrium. For  $L>0$, we also have to take the continuum contribution
into account. In has been shown \cite{Wendin} that the
continuum contribution to the equilibrium current is
$I_{eq}^c=-1/2I_{eq}^b$, thus giving the total equilibrium current
$I_{eq}=1/2I_{eq}^b$.

At zero temperature, when a voltage equal to the lowest lying level
$eV=E_0^-$ is applied, the regular and anomalous current jumps

\begin{equation}
\delta
I_r=\frac{e\Delta}{\hbar}\frac{L}{\xi_0}\frac{\sqrt{D}\sin(\phi)}{2|\sin(\phi/2)
|\sqrt{1-D\sin^2(\phi/2)}},
\end{equation}

\begin{equation}
\delta
I_a=\sigma \frac{e \Delta}{\hbar}\frac{L}{\xi_0}
\frac{\sqrt{RD}\sin\phi}{\sqrt{1-D\sin^2(\phi/2)}}.
\end{equation}
Both jumps are proportional to $\sqrt{D}$, and the magnitude of the total
current at $E_0^-<eV<E_0^+$ is then much larger than the equilibrium
current.

When the bias voltage is further increased to $eV=E_0^+$ there is a second
current jump: the regular current jumps in the {\em opposite} direction and
becomes equal to the small negative bound state equilibrium current
$I_r=-I_{eq}^b$. The anomalous current, however, again jumps $\delta I_a$
in the {\em same} direction. For voltages $eV>E_0^+$ the total current in
the junction is thus $I_{eq}^c+2\delta I_a$. The full formulas for all the
individual currents including temperature dependence is given in Appendix C.

\section{Josephson current of a long junction}

We now discuss the Josephson current in a long ($L\gg\xi_0$) symmetric
($l=0$) junction, and we treat the effects of asymmetry below.  In a
long junction there are many $(N=[L/(\xi_0 \pi)])$ pairs of
resonances, as seen in Fig. \ref{asandr}. The width of each resonance
is $\Gamma=\epsilon \hbar v_F/(2L)$. This width must not be too small
if the quasiparticles are to be able to enter and leave the junction
without being scattered inelastically ($\Gamma > \hbar
v_F/l_i$). This gives an upper limit for the length $L < l_i
\epsilon$.

The derivative of energy with respect to phase $dE/d\phi$ in Eq.
(\ref{asymdedfi}), which determines the current in Eqs. (\ref{irbound})-
(\ref{sumboundstate}), can be simplified in a long junction
$L\gg\xi_0$,

\begin{equation}
\frac{dE^{\pm}}{d\phi}=\pm\frac{\hbar v_F}{L}
\frac{\sqrt{D}\sin(\phi)}{4|\sin(\phi/2)|\sqrt{1-D \sin^2(\phi/2)}}.
\label{longdedfi}
\end{equation}
This expression holds everywhere except close to the gap edge,
$\Delta-E_n \sim (\hbar v_F/L)(\xi_0/L)$, a distance much smaller than
the energy distance $\pi\hbar v_F/L$ between the pairs of
levels. Therefore, equation (\ref{longdedfi}) can be used for
calculation of the currents of all levels except the last pair of
levels closest to the energy gap. The current from this last pair of
levels must always be treated on a separate footing.

According to Eq. (\ref{longdedfi}), each of the Andreev levels carry
a current of the order of $1/L$. Furthermore, as follows from the
exact Eq. (\ref{asymdedfi}), each pair of levels carries a small
net current, $dE_{n}^+/d\phi-dE_{n}^-/d\phi$, of the order of
$(1/L)^3$. The sum of the currents from all bound states is thus
determined by the current ($\sim 1/L$) from the last pair of levels.

\subsection{Equilibrium current}

For the equilibrium current of a long junction, the contribution from
the Andreev bound states at $|E|<\Delta$ and from the continuum at
$|E|>\Delta$ are of the same magnitude (see
Fig. \ref{boundcontcurr}). The continuum current (see Appendix A) is
given by

\begin{eqnarray}
&&I_{eq}^{c}=\frac{e}{h}\left(\int_{-\infty}^{-\Delta}+\int^{\infty}_{\Delta}\right)
dE n_F\nonumber \\
&&\times \frac{4D\sin\phi\sin\beta \sinh 2\gamma_c
}{(\cos\beta\cosh2\gamma_c-R-D\cos\phi)^2+(\sin\beta \sinh 2
\gamma_c)^2},
\label{contcurr2}
\end{eqnarray}
with $\gamma_c=\mbox{arccosh}(E/\Delta)$. Following the method by
Ishii \cite{Ishii} and Svidzinsky et al. \cite{Kulik}, one can
rewrite this integral as a sum over the residues,

\begin{equation}
I_{eq}^{c}=-I_{eq}^{b}+4kT\pi i\sum_{p=0}^{\infty}i(E_p),
\label{contcurr3}
\end{equation}
where the first term results from the poles of the current density
$i(E)$ in Eq. (\ref{contcurr2}) and the second term from the poles of
the distribution function $n_F(E)$, given by $E_p=i2kT\pi(1/2+p)$. The
first term in (\ref{contcurr3}) is the current carried by the bound
states with negative sign. The total equilibrium current $I_{eq}$ is
then just given by the second term.  For zero temperature this sum
over poles turns into an integral and the total equilibrium current is
then given by

\begin{equation}
I_{eq}(T=0)=\frac{e}{\hbar}\frac{\hbar
v_F}{L}\frac{\sqrt{D}\sin(\phi)\arccos(R+D\cos\phi)}{2\pi
|\sin(\phi/2)|\sqrt{1-D \sin^2(\phi/2)}}.
\label{longzerotemp}
\end{equation}
%hg
At high temperatures $\hbar v_F/L\ll kT \ll \Delta$, only the the first term
$(p=0)$ in the sum $(\ref{contcurr3})$ needs to be included, and the
equilibrium current becomes

\begin{equation}
I_{eq}(kT \gg \hbar v_F/L)=\frac{e \Delta}{\hbar}\pi
\left(\frac{4kT}{\Delta}\right)^2 D \sin \phi e^{-2\pi L/\xi_T},
\label{longfinitetemp}
\end{equation}
where $\xi_T=\hbar v_F/kT$. The equilibrium current for a
long junction at finite temperature is thus exponentially small.
\cite{Kulik}. Expressions (\ref{longzerotemp}) and
(\ref{longfinitetemp}) extend earlier results \cite{Kulik,Ishii}
to the case of arbitrary transparency $D$ of the junction.

\begin{figure}
\centerline{\psfig{figure=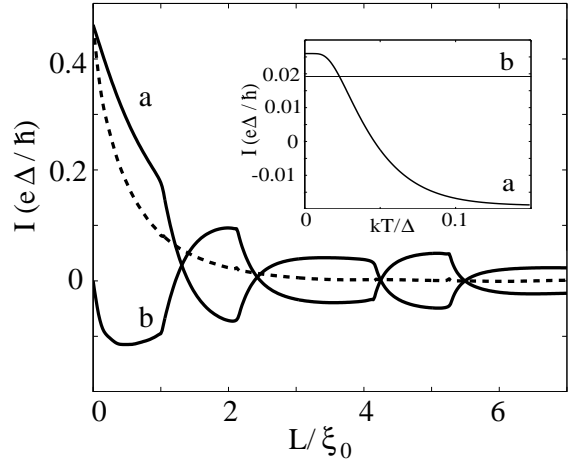,height=6.5cm}}
\caption{The equilibrium bound state (a) and continuum (b) currents
and their sum (dashed) as a function of length $L$ for finite
$kT=0.2\Delta$, $D=0.8$, $\phi=3\pi/4$ and $\epsilon=0.01$. There is a
cusp in both the bound state and continuum currents when a new bound
state forms out of the continuum. The total equilibrium current,
however, dies monotonically with increased length. Inset: The
equilibrium bound state and continuum currents as a function of
temperature for a long junction $L=15\xi_0$ with $D=0.8$, $\phi=3
\pi/4$ and $\epsilon=0.01$. The bound state current (a) decreases from
$I_{eq}^b(T=0)=i^*$ to $-I_{eq}^c$, when the temperature is increased
from zero to $kT \ll \hbar v_F/L$. The continuum current (b) is
unaffected in this temperature regime.}
\label{boundcontcurr}
\end{figure}

We are also interested in analyzing the separate behavior of the
bound state current, because this current is revealed in
nonequilibrium, as will be discussed in detail below. Therefore, using
relation (\ref{longdedfi}) we can write Eq. (\ref{sumboundstate}) on
the form

\begin{eqnarray}
I_{eq}^{b}&=&\frac{e}{\hbar}\frac{\hbar
v_F}{L}\frac{\sqrt{D}\sin(\phi)}{2|\sin(\phi/2)|\sqrt{1-D\sin^2(\phi/2)}}
\nonumber \\
&&\times \sum_{n=0}^{N-1} \left[\tanh(E_n^-/2kT)-\tanh(E_n^+/2kT)\right]
\nonumber \\
&& +i^{*}\tanh(\Delta/2kT).
\label{longeq}
\end{eqnarray}
The term $i^*$ results from the last pair of levels at
$E\approx\Delta$, and is of the order $1/L$, as discussed above. At
$T=0$, the sum in Eq. (\ref{longeq}) is zero, and we thus find that
$i^{*}=I_{eq}^b(T=0)$. When the temperature is increased, the sum in
Eq. (\ref{longeq}) starts to contribute with negative sign and the
bound state current is decreased. The continuum current (and also
$i^*$), however, is independent of temperature for $kT\ll\Delta$,
since it is an integral over states with $|E|>\Delta$ [see
Eq. (\ref{contcurr2})]. At $kT \gg \hbar v_F/L$ the {\em total}
equilibrium current is exponentially small (see
Eq. (\ref{longfinitetemp}) and has thus decreased an amount
$I_{eq}(T=0)-I_{eq}(kT\gg\hbar v_F/L) \approx I_{eq}(T=0)$. This is
thus solely due to decrease of the bound state current, as shown in
the inset in Fig. \ref{boundcontcurr}.

\subsection{Regular current}

\begin{figure}
\centerline{\psfig{figure=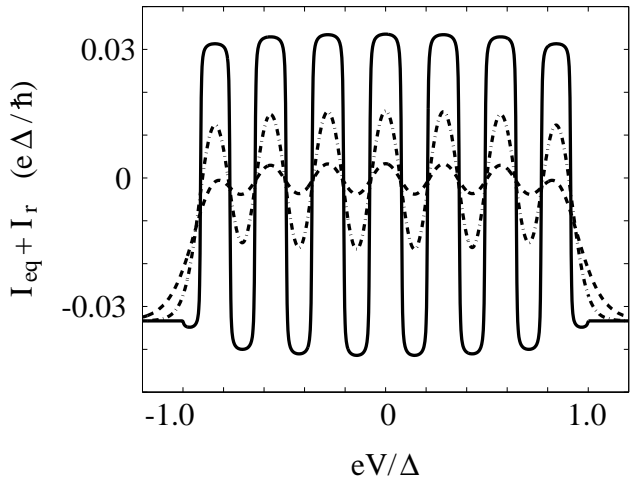,height=6.5cm}}
\caption{The equilibrium current $I_{eq}$ plus the regular current $I_r$ vs
voltage. $L=10 \xi_o$, $\phi=\pi/2$, $D=0.8$, $\epsilon=0.05$.  Solid
line - $T=0$, dashed-dotted - $kT=0.04\Delta$, dashed -
$kT=0.07\Delta$. The regular current jumps alternating by $\pm \delta I_r$
every time the voltage is equal to the energy of an Andreev
resonance. For $kT \gg \hbar v_F/L$ and $eV>\Delta$ the current
$I_{eq}+I_r=I_{eq}^c$}
\label{ireg}
\end{figure}

The regular current can be written, inserting relation
(\ref{longdedfi}) into Eq. (\ref{irbound}), on the form

\begin{eqnarray}
I_r&=&\frac{e}{\hbar}\frac{\hbar
v_F}{2L}\frac{\sqrt{D}\sin(\phi)}{2|\sin(\phi/2)|\sqrt{1-D\sin^2(\phi/2)}}
\nonumber \\
&& \times \sum_{n=0}^{N-1}[g(E_n^-)-g(E_n^+)]+\frac{i^*}{2}g(\Delta)
\label{ireglong}
\end{eqnarray}
where $g(E)=\tanh[(E+eV)/2kT]+\tanh[(E-eV)/2kT]-2\tanh(E/2kT)$. The regular
current $I_r$ jumps up or down every time $eV=E_n^{\pm}$ [see Fig.
\ref{ireg}]. Each current jump has the magnitude

\begin{equation}
\delta I_r=\frac{e}{\hbar}\frac{\hbar
v_F}{L}\frac{\sqrt{D}\sin(\phi)}{2|\sin(\phi/2)|\sqrt{1-D\sin^2(\phi/2)}}
\end{equation}
at zero temperature. At voltages $eV>\Delta$, the regular current is
the sum of all states in the range $0<E<\Delta$, and is equal to
the negative bound state current $-i^*$.

It is interesting to study the sum $I_{eq}+I_r$, plotted in
Fig. \ref{ireg}, at temperatures $kT\gg \hbar v_F/L$. In this
temperature regime the equilibrium current is exponentially small and
also the regular current steps in the IVC in Fig. \ref{ireg} are
suppressed. For a voltage $eV\sim \Delta$, the last level, carrying the
major part ($i^*$) of the bound state current, is populated and the
current $I_{eq}+I_r$ jumps to $I_{eq}^c$, the value of the continuum
current, since all bound states are populated. This current $I^c_{eq}$
is of the order of $1/L$ and the current $I_r+I_{eq}$ is {\it
increased } from zero to be $\sim 1/L$ when increasing
the voltage from $V=0$ to $eV \sim \Delta$.

\subsection{Anomalous current}

\begin{figure}
\centerline{\psfig{figure=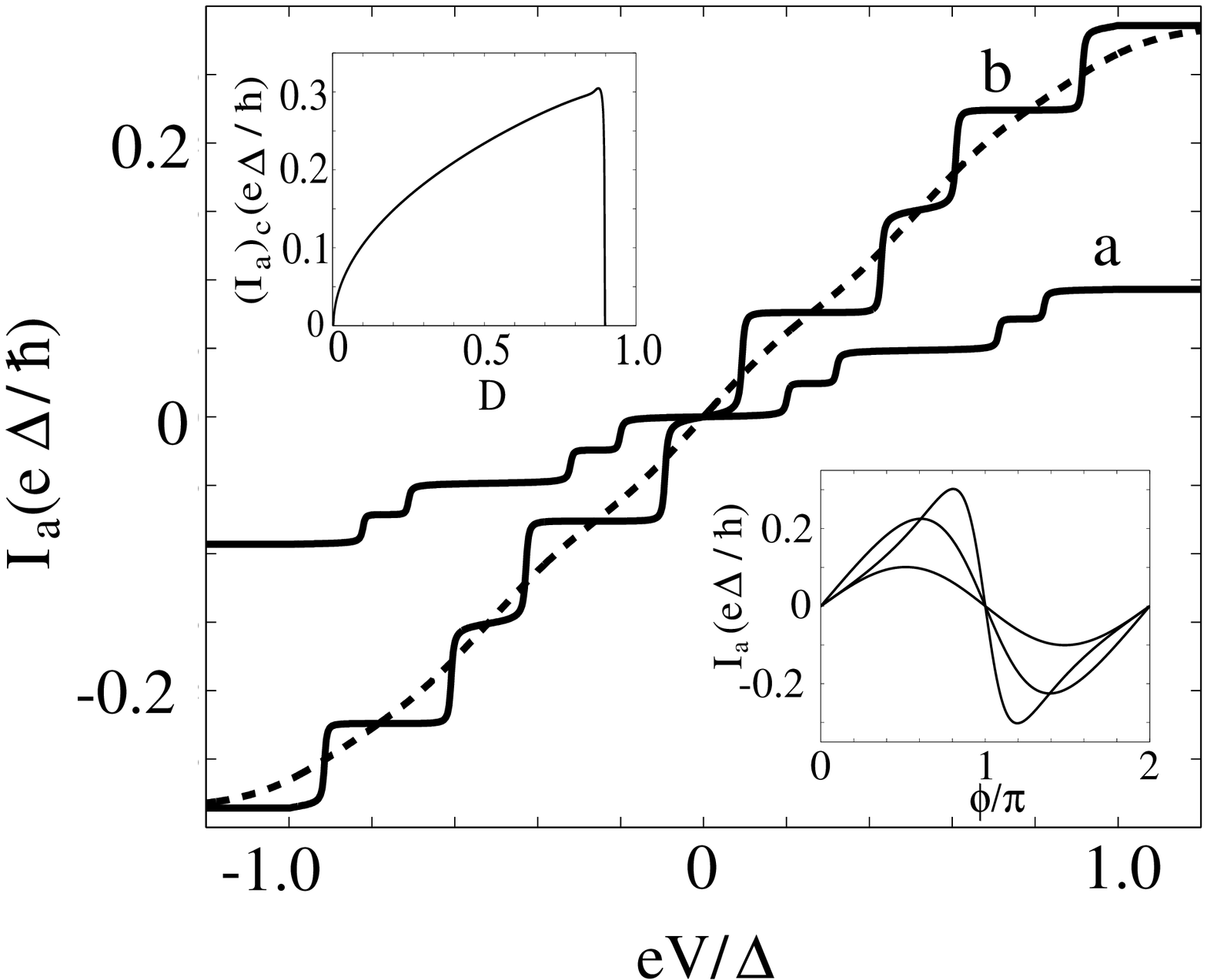,height=6.5cm}}
\caption{The anomalous current as a function of voltage $V$ for (a)
$\phi=\pi/4$ and (b) $\phi=3\pi/4$ for $L=10 \xi_0$, $D=0.8$,
$\epsilon=0.05$ and $\sigma=-1$. Temperature $T=0$ (solid) and
$T=0.1\Delta$ (dashed line). The current steps with magnitude $\delta
I_a$ are smeared to a straight line for $kT \gg \hbar v_F/L$. Upper inset: The
critical anomalous current at $eV=\Delta$ as a function of
transparency $D$ for coupling constant $\epsilon=0.1$. Due to finite
coupling $\epsilon$, the critical current always goes to zero for
$R=0$. Lower inset: The anomalous current $I_a(eV=\Delta,kT \gg \hbar v_F/L)$
as a function of phase difference $\phi$ for different transparencies
$D=0.1, 0.5$ and $0.9$. The highest amplitude corresponds to the
highest transparency and vice versa.}
\label{panvolt}
\end{figure}

The anomalous current is given by inserting Eq.
(\ref{longdedfi}) into Eq. (\ref{iabound}),

\begin{equation}
I_a=-\sigma\frac{e}{\hbar}\frac{\hbar
v_F}{4L}\frac{\sqrt{RD}\sin\phi}{1-D\sin^2(\phi)}\sum_{n=0}^{N-1}[h(E_n^+)+h(E
_n^-)],
\label{longian}
\end{equation}
where $h(E)=\tanh[(E-eV)/2kT]-\tanh[(E+eV)/2kT]$. We have neglected
the current from the last level close to $E=\Delta$, because the
currents of all levels add up and the current from the last level is
negligible. The IVC at zero temperature looks like a staircase, as
shown in Fig.  \ref{panvolt}.

The magnitude of the current step at zero temperature is given by

\begin{equation}
\delta I_a=\frac{e}{\hbar}\frac{\hbar
v_F}{2L}\frac{\sqrt{RD}\sin\phi}{1-D\sin^2(\phi/2)}.
\end{equation}
At temperatures larger then the interlevel distance, $kT \gg \hbar
v_F/L$, the staircase IVC is smeared out to a straight slope, as shown
in Fig. \ref{panvolt}. The exact position of each level becomes
irrelevant and we can write the sum over bound states in
(\ref{longian}) as an integral, noting that the expression
$dE/dn=\pi\hbar v_F/L$ holds for all levels in the sum
(\ref{longian}),

\begin{eqnarray}
&&\sum_{n=0}^{N}[h(E_n^+)+h(E_n^-)] \nonumber \\
&&\approx \frac{2L}{\pi \hbar
v_F}\int_{0}^{\Delta}dE[\tanh(E+eV)-\tanh(E-eV)] \nonumber \\
&&=\frac{4L}{\hbar v_F \pi}f(V,T)
\label{sumint}
\end{eqnarray}
where
\begin{equation}
f(V,T)=kT\ln\left(\frac{\cosh(\Delta+eV)/kT}{\cosh(\Delta-eV)/kT}\right),
\label{fdef}
\end{equation}
and the anomalous current takes the simple form

\begin{equation}
I_a=-\sigma\frac{e}{\hbar}\frac{\sqrt{RD}\sin\phi}{\pi[1-D\sin^2(\phi/2)]}f(V,T)
.
\label{iasimple}
\end{equation}
In the limit $\hbar v_F/L\ll kT\ll \Delta$, $f(V,T)=\min(eV,\Delta)$:
the anomalous current thus scales linearly with applied voltage up to
$\Delta$. It follows from Eq. (\ref{iasimple}) that $I_a$ is
independent of the length of the junction, being the sum of $N\sim L$
levels which each carries a current $I_n\sim 1/L$. This gives that the
anomalous current roughly is equal to the total equilibrium current of
the short junction. The critical anomalous current is plotted with
respect to transparency in the inset in Fig.~\ref{panvolt}. In the
limit $D\ll 1$ it is given by
$(I_a)_c=(e/\hbar)(\sqrt{D}/\pi)f(V,T)$. It is proportional to the
first power of $\Delta$ for $T$ close to $T_c$, therefore surviving up
to $kT\approx\Delta$. The anomalous current-phase relation (see inset
in Fig. \ref{panvolt}) is $2\pi$ periodic and resembles that of the
equilibrium Josephson current.  The direction of the anomalous current
is however proportional to $\sigma$, i.e dependent on the phase of the
scatterer at the connection point, which is not the case for the
equilibrium Josephson current.

\begin{figure}
\centerline{\psfig{figure=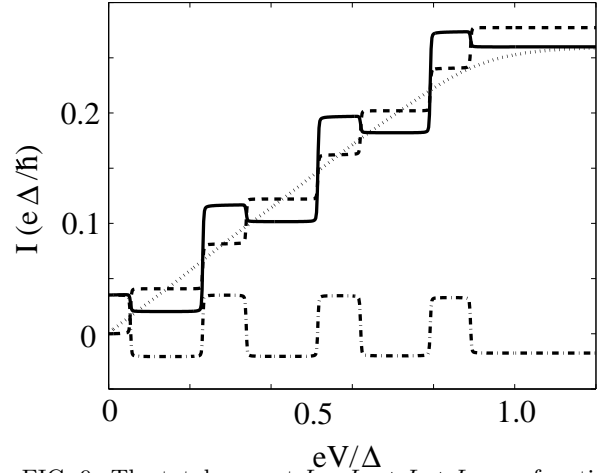,height=6.5cm}}
\caption{The total current $I=I_{eq}+I_r+I_a$ as a function of
voltage. At zero temperature we have $I_r+I_{eq}$ (dash-dotted), $I_a$
(dashed) and the total current $I_r+I_{eq}+I_a$ (solid). The total
current for temperatures $kT \gg \hbar v_F/L$ is plotted
(dotted). Junction parameters are $D=0.8, \phi=3\pi/4, L=20 \xi_0$,
$\epsilon=0.01$ and $\sigma=-1$}
\label{totalcurr}
\end{figure}

To get the complete picture of the Josephson current in a long
junction, $I=I_{eq}+I_r+I_a$ is plotted as a function of bias voltage
for different temperatures in Fig. \ref{totalcurr}.

The zero temperature total Josephson current oscillates strongly
around a constant slope as a function of voltage, showing steps
whenever the voltage passes an Andreev level. The step structures are
washed out for temperatures $kT \gg \hbar v_F/L$, and in this limit
the total current roughly coincides with the anomalous current, given
by Eq. (\ref{iasimple}).

\subsection{Asymmetric junction}

The effect of asymmetry is most pronounced in the long limit when the
asymmetry is much larger than the coherence length but much smaller
than the total length of the junction, $L \gg l \gg \xi_0$. In this
limit, the derivative of energy with respect to phase $dE/d\phi$ in
Eq. \ref{asymdedfi} reduces to the expression of a symmetric long
junction (\ref{longdedfi}), since $|\sin 2\theta|>R|\sin \chi|$ (see
Appendix B). The equilibrium current $I_{eq}$ and the regular current
$I_r$ are not substantially changed in comparison to the symmetric
case. In contrast, the anomalous current is modified in a nontrivial
way, taking the form

\begin{eqnarray}
I_a&=&-\sigma {e\over \hbar} \frac{\hbar v_F}{L} \sqrt{R}D^{3/2}
\sin\phi \nonumber \\ &&\times
\sum_{n,\pm}^N\frac{\cos\chi-\cos\phi}{1-(D\cos\phi+R\cos\chi)
^2}(n^e-n^h),
\label{ianascurr}
\end{eqnarray}
obtained by inserting Eq. (\ref{longdedfi}) into (\ref{iabound}).
For $T=0$ the step structure in the IVC is modified due to the change
of Andreev levels as a result of the asymmetry (see
Fig.~\ref{asandr}). Already for small asymmetry $l\sim\xi_0$ the
anomalous current might change dramatically. Depending on the phase
difference of the junction, the IVC is renormalized and changes sign
for $-\pi/2<\phi<\pi/2$.

\begin{figure}
\centerline{\psfig{figure=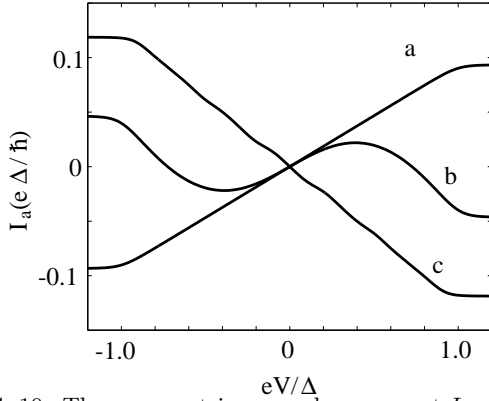,height=5.5cm}}
\caption{The asymmetric anomalous current $I_{a}$ vs voltage for
different asymmetries (a) $l=0\xi_0$, (b) $l=2\xi_0$ and (c)
$l=40\xi_0$ for $kT\gg \hbar v_F/L$, $D=0.8$, $\epsilon=0.05$,
$L=60\xi_0$, $\phi=\pi/4$ and $\sigma=-1$. The IVC is changed
dramatically already for as small asymmetry $l\sim \xi_0$, if the
phase difference $-\pi/2<\phi<\pi/2$}
\label{ianas}
\end{figure}
When the temperature is increased beyond the interlevel distance
$kT\gg \hbar v_F/L$, the step structure becomes smeared and we get a
periodic modulation of the IVC on the scale of $eV\sim\hbar
v_F/l$. This modulation arises from the factor $\cos \chi$.

\begin{figure}
\centerline{\psfig{figure=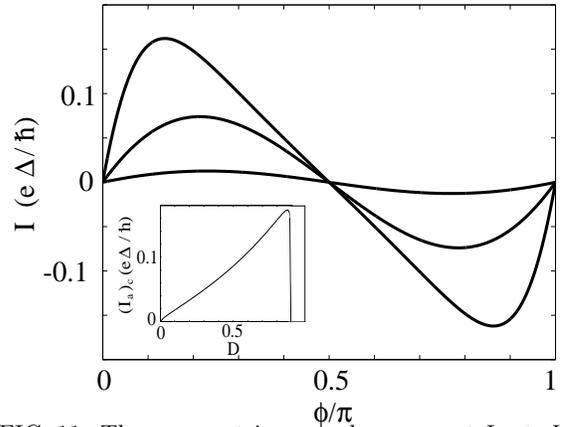,height=6.5cm}}
\caption{The asymmetric anomalous current $I_a$ at $eV=\Delta$ and $kT\gg
\hbar v_F/L$ as a function of phase difference $\phi$ for different
transparencies $D=0.1, 0.5$ and $0.9$. Inset: The critical anomalous
current $(I_a)_c$, for $eV=\Delta$ and $kT\gg \hbar v_F/L$, as a function of
transparency $D$ for coupling constant $\epsilon=0.1$.}
\label{ianasph}
\end{figure}

When the temperature is further increased to $kT\gg \hbar v_F/l$ this
periodic structure is smeared out and the IVC once again becomes a
straight line, but with renormalized slope. In this high temperature limit
the amplitude of the terms in the sum in Eq. (\ref{ianascurr}) oscillates
with a period $\hbar v_F/l$. During this period, the filling factors
$n$ can be taken to be constant, and we can sum over one period to get the
average value. Performing this summation in the continuum limit, we get

\begin{eqnarray}
&&\sum_{one~period}\frac{\cos\chi-\cos\phi}{1-(D\cos\phi+R\cos
\chi)^2} \nonumber \\ &&\approx \frac{\hbar
v_F}{2l}\int_0^{2\pi}\frac{\cos(\chi)-\cos(\phi)}{1-(D\cos\phi+R
\cos\chi)^2} d\chi \nonumber \\
&&=\frac{L}{l}\frac{1}{8R\sqrt{D}}\left(
\frac{|\sin(\phi/2)|}{\sqrt{1-D\cos^2(\phi/2)}}-\frac{|\cos(\phi/2)|}{\sqrt{1-D\sin^2(\phi/2)}}\right). \nonumber \\
\label{ianassum}
\end{eqnarray}
This quantity is energy independent and we can then sum over the filling
factors
following the procedure from the symmetric case (\ref{sumint})

\begin{equation}
\sum_{averaged~periods}(n^e-n^h)\approx\frac{4l}{\hbar v_F\pi}f(V,T).
\end{equation}
The anomalous current then becomes

\begin{eqnarray}
&&I_a=-\sigma 2 {e\over \hbar} \frac{D}{\pi \sqrt{R}} \sin\phi
\nonumber \\
&&\times
\left(\frac{|\sin(\phi/2)|}{\sqrt{1-D\cos^2(\phi/2)}}-\frac{|\cos(\phi/2)|}{\sqrt{1-D\sin^2(\phi/2)}}\right)f(V,T), \nonumber \\
\label{renormanasym}
\end{eqnarray}
which is independent of both the length $L$ and the asymmetry $l$. We
also find that the renormalized anomalous current becomes
$\pi$-periodic. This can qualitatively be explained by the fact that
the $2\pi$-periodic part of the anomalous current density is very
sensitive to asymmetry, oscillating fast with energy on the scale of
$\hbar v_F/l$, becoming washed out during summation over bound states
at high temperatures $kT\gg\hbar v_F/l$. The $\pi$ periodic part of
the current does not have this sensitivity and is the only part of the
anomalous current that survives. The asymmetric anomalous
current-phase relation is shown in Fig. \ref{ianasph}.

The $\pi$-periodicity and the zeros at $\phi=n\pi/2$ give the
condition that the slope of the IVC must change sign due to asymmetry
in the range $-\pi/2<\phi<\pi/2$, as shown in Fig. \ref{ianas}. The
critical asymmetric anomalous current as a function of transparency
$D$ is shown in the inset in in Fig. \ref{ianasph}. The critical
asymmetric anomalous current as a function of transparency $D$ is
shown in the inset in in Fig. \ref{ianasph}. The behavior is very
similar to the critical anomalous current in the symmetric case, the
main difference being that the amplitude is reduced by roughly a
factor of two.

\section{Interface barriers}

In any realistic experimental situation, normal reflection at the
NS-interface, modeled by a barrier with reflection amplitude $r_b$,
must be taken into account. \cite{Chrestin} The
general expression, considering both the interface barriers and the
midpoint scatterer, becomes analytically intractable. We can however
analyze the case where the midpoint scatterer is absent ($R=0$) to get
an understanding of the effect of NS-barriers on the junction
properties, and then treat the general case with injection and
midpoint scatterer numerically.

\begin{figure}
\centerline{\psfig{figure=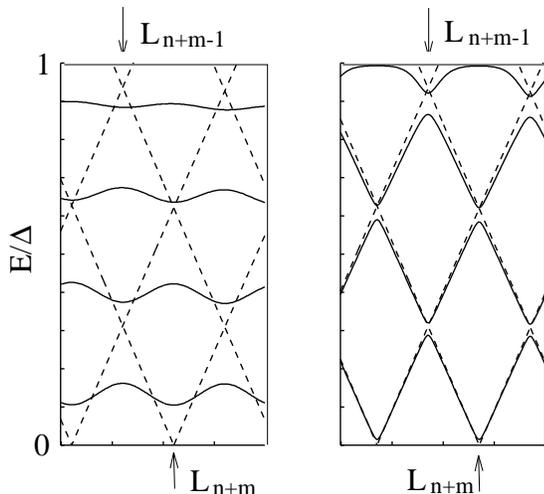,height=6.5cm}}
\caption{The Andreev levels (solid) and the normal electron and hole
resonances (dotted) as a function of length $L$ of the junction with
four Andreev levels in (a) weak resonance limit $R_b\ll 1$ (b) strong
resonance limit $R_b\sim 1$. The lengths of two subsequent
intersections of normal resonances $L_{n+m}$ and $L_{n+m-1}$ are shown
with arrows}
\label{bwandrres}
\end{figure}

In the absence of the superconducting leads (a NININ-junction), the
two barriers give rise to normal Breit-Wigner resonances for the
electrons and holes. Understating the properties of these resonances
turns out to be crucial for describing the behavior of Andreev levels
and current transport. The energies of the electron and hole
resonances are calculated straightforwardly

\begin{eqnarray}
E^e_n=-2E_F\left[1-\frac{\pi(n-\nu)}{k_FL}\right], \nonumber \\
E^h_m=2E_F\left[1-\frac{\pi(m-\nu)}{k_FL}\right],
\end{eqnarray}
where $r_b=\sqrt{R_b}e^{i\nu\pi}$ and $n(m)$ are integers denoting the
index of the electron (hole) resonances. The intersection between
electron and hole resonances ($E^e_n=E^h_m$) is given by
$L_{n+m}=\lambda_F/4(m+n-2\nu)$ with the Fermi wavelength
$\lambda_F=2\pi/k_F$. These normal resonances are plotted in
Fig. \ref{bwandrres}.

For the junction with superconducting leads, one can in the same way
as before calculate the equation for the bound Andreev states
($|E|<\Delta$), with
the result \cite{Wendin2}

\begin{eqnarray}
&&D_b^2\cos\phi+2R_b\cos\beta-\cos(2\gamma-\beta)-R_b^2\cos(2\gamma+\beta)
\nonumber \\
&&-4R_b\sin^2\gamma\cos(\beta_0)=0
\end{eqnarray}
where we have defined $\beta_0=\pm 2E^{e,h}/(\hbar v_F/L)$ and $+(-)$ denotes
hole(electron) resonance energies. One can draw some qualitative conclusions
on how the Andreev levels are related to the normal resonances by looking
at Fig. \ref{bwandrres}. In the limit of high barrier transparency
$R_b\ll 1$, the Andreev levels are
weakly modified by the barriers. In the opposite limit $R_b\sim 1$, the
Andreev levels get pinned at the normal resonances, but there are no
level crossings at the points where the normal electron and hole
resonances intersect.

We find the same interlevel distance $\hbar v_F \pi/L$ in the junction
with the superconducting leads (SINIS junction) and normal leads
(NININ-junction). The main difference is that the normal resonance
move very quickly through the junction when the length $L$ increases,
while the Andreev levels oscillate up and down.

Considering Andreev state energies close to the Fermi level,
$E\ll \Delta$, one can derive a simplified dispersion relation
\cite{Kadigrobov2}

\begin{equation}
\sin^2(\beta/2)=\frac{D_b^2\cos^2(\phi/2)+4R_b\sin^2(\beta_0/2)}{(1+R_b)^2}.
\label{barrel}
\end{equation}
Using this relation we can study the bound state current in different
length limits.

In the short limit, $L\ll \xi_0$ there are two cases to be considered. For
nearly transparent barriers $D_b\sim 1$, and thus broad resonances
$\Gamma=D_b
\hbar v_F/L\gg \Delta$, one can neglect dephasing (putting $\beta=0$) and
just get the total transparency of the junction
$D=D_b^2/(D_b^2+4R_b\sin^2(\beta_0/2))$ to be put into the standard zero
length junction equilibrium current formula.  In the strong barrier case
$D_b\ll 1$ the resonances are sharp $\Gamma\ll \Delta$ and one can not
neglect the dephasing. Assuming that the resonance is close to Fermi energy
$E^{e,h}\ll\Delta$, we can put $\beta\ll 1$ in (\ref{barrel}) and
obtain\cite{Wendin2,Beenakker4}

\begin{equation}
E=\pm \sqrt{\Gamma^2\cos^2(\phi/2)+(E^{e,h})^2}.
\end{equation}
When the resonance is exactly at the Fermi energy $E^{e,h}=0$, the
Josephson current is given by

\begin{equation}
I=\frac{e\Gamma}{\hbar}\sin(\phi/2)\tanh\left(\frac{\Gamma\cos(\phi/2)}{2kT}
\right).
\end{equation}
The critical current at low temperatures ($kT\ll \Gamma$) is thus smaller
than the critical current of a short, clean junction by a factor
$\Gamma/\Delta$.

For a long junction $L\gg\xi_0$ we can calculate the derivative of energy
with respect to phase,

\begin{equation}
\frac{dE}{d\phi}=\pm\frac{\hbar
v_F}{2L}\frac{D_b^2\sin\phi}{\sqrt{(1+R_b)^4-[D_b\cos\phi-4R_b\cos(\beta_0)]^2}}
,
\label{dedfibarr}
\end{equation}
In the weak barrier limit $R_b\ll 1$, this just causes oscillations
with length around the clean junction ($R_b=0$) result. In the strong
barrier limit $R_b\sim 1$, one can distinguish two limits: When the
length of the junction is far away from the length $L_{n+m}$ where the
electron and hole resonances intersect, the junction is out of
resonance. The second term in Eq. (\ref{dedfibarr}) is negligible and
the current from the individual levels thus becomes

\begin{equation}
I=\pm\frac{e v_F}{4L} D_b^2\sin\phi.
\end{equation}
It is proportional to $D_b^2$ and thus strongly suppressed. In the
opposite limit, when the length of the junction
$L=L_{n+m}=\lambda_F(m+n-2\nu)/4$, the junction is in resonance. When
$n+m$ is even we get the current carried by each level

\begin{equation}
I=\pm\frac{e v_F}{L}\frac{D_b\sin\phi}{4|\cos(\phi/2)|}
\label{ires1}
\end{equation}
and when $n+m$ is odd we get

\begin{equation}
I=\pm\frac{e v_F}{L}\frac{D_b\sin\phi}{4|\sin(\phi/2)|}.
\label{ires2}
\end{equation}
We see that the current is proportional to $D_b$, just as expected for
the junction in resonance. The current carried is thus of the order of
the single barrier junction current. An interesting feature is that
the current is dependent on the parity of the sum of the electron and
hole resonance indices $n+m$. When the third lead is connected to the
junction, the scattering at the connection point just splits the Breit
Wigner resonances, and the qualitative picture for the bound states
derived without the third lead connected survives.

To calculate the total equilibrium, regular or anomalous current, the
currents carried by all individual levels have to be summed up. In the
weak barrier limit $R_b\ll 1$ we just find that all properties
calculated above for the symmetric junction without barriers hold,
with a small length dependent modulation $\sim R_b$ with a period
$\delta L\sim \lambda_F$. In the strong barrier limit $R_b\sim 1$, the
result will depend on whether the junction is in or out of resonance.

\begin{figure}
\centerline{\psfig{figure=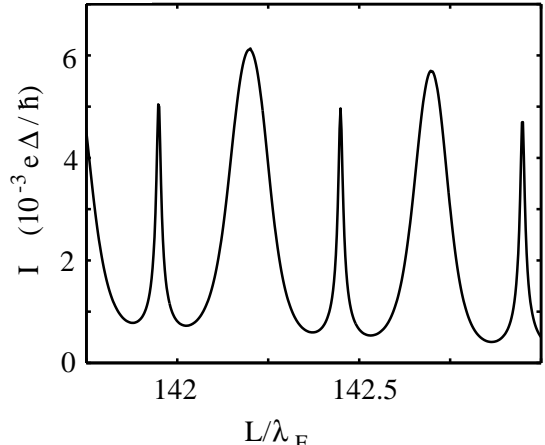,height=6.5cm}}
\caption{Short segment of the equilibrium bound state current as a
function of length $L$, to illustrate the resonant behavior. The
junction is long $L\gg \xi_0$ with $R_b=0.9$, $\epsilon=0.01$ and
$\phi=\pi/2$.}
\label{paseq}
\end{figure}
Fig. \ref{paseq} shows the resonant behavior of the equilibrium
current as a function of length. The current has a peak around lengths
$L=\lambda_F/4(m+n+2\nu)$. The phase dependence of the current at the
resonant peaks is well described by the expressions for the single
level currents (\ref{ires1}) and (\ref{ires2}).

The anomalous current is also strongly length dependent and when the
junction is in resonance we have an anomalous current $I_a\sim \sigma
D_b \sqrt{RD}$ while when we are out of resonance $I_a\sim \sigma
D_b^2 \sqrt{RD}$. It turns out that there is an anomalous current even
without scattering at the connection point, but it oscillates around
zero as a function of length with the period $\sim \lambda_F$.

\section{Injection current and conductance}

Although the nonequilibrium Josephson current is at the focus of this
paper, the injection current that flows between the normal reservoir
and the SNS-junction is also of great interest: it determines the
conductance of the circuit. The conductance of NS-structures has been
studied intensely in recent years \cite{Lambertrew} and is an
interesting quantity in itself. It can also be used to determine the
direction of the Josephson current in the junction \cite{Vanwees} or
to detect a large Josephson current in the superconducting loop which
changes the applied external flux vs phase dependence, thus modifying
the phase dependence of the conductance. \cite{Phaserel}

\subsection{Injection current}

\begin{figure}
\centerline{\psfig{figure=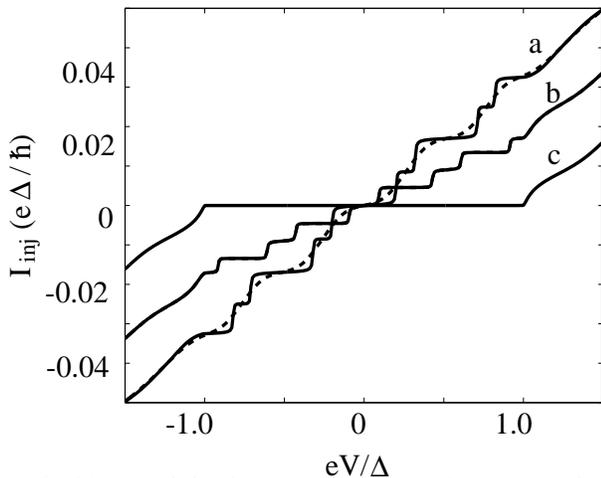,height=6.5cm}}
\caption{The injection current in lead 1 as a function of voltage for
(a) $\phi=\pi/4$, (b) $\phi=3\pi/4$ and (c) $\phi=\pi$. Zero $T$
(solid lines) and $T=0.05\Delta$ (dashed line) with $D=0.8$,
$L=10\xi_0$ and $\epsilon=0.05$. For $eV>\Delta$ the IVC approaches
the value of a normal junction}
\label{injection}
\end{figure}

We start by discussing the symmetric junction $l=0$, and comment on
the modifications due to asymmetry below.  The current injected in
lead $1$ for energies $|E|<\Delta$ can be calculated to lowest order
in $\epsilon$ by inserting Eq. (\ref{injcurrdens}) into
Eq. (\ref{i1def}) 

\begin{eqnarray}
I_1&=&\epsilon\frac{e}{\hbar}\frac{|\cos(\phi/2)|}{\sqrt{D}\sqrt{1-D\sin^2(\phi/
2)}}
\nonumber \\ &&\times
\sum_{n,\pm}\left|\frac{dE_n^{\pm}}{d\phi}\right|\left[n^e(E_n^{\pm})-n^h(E_n^{\pm})\right].
\label{iinj}
\end{eqnarray}
This current is proportional to $\epsilon$ (unlike the Josephson
current discussed above), which is also true for the continuum
contribution. For $|E|>\Delta$, the injection current density in
Eq. $(\ref{i1def})$, $i_{inj} = i_1^-=i_3^--i_2^-$, is obtained from
Eq. (\ref{contdiffcurr}). This current density is roughly described by
the normal current density $4\epsilon e/h$, with oscillations around
this value due to the resonances in the scattering states (see
Fig. \ref{injection}). These oscillations are strongest around $E\sim
\Delta$ and decrease with increasing energy.

The injection current is proportional to the modulus $|dE/d\phi|$,
just like the anomalous current, as discussed above. The IVC thus has
the shape of a staircase, as shown in Fig. \ref{injection}.

We can derive expressions for the injection current in different
length limits. A complete set of formulas are given in Appendix A.  Here we
only present the heights of the current steps in the IVC:s in some
representative cases.

In the limit of zero length of the junction ($L=0$) there is only one
Andreev level (for $0<E<\Delta$) and the magnitude of the step is

\begin{equation}
\delta
I_1=\epsilon\frac{e}{\hbar}\frac{\sqrt{D}}{2}\frac{|\sin(\phi/2)|\cos^2(\phi/2)}
{1-D\sin^2(\phi/2)}
\end{equation}
while for two levels close to $\Delta$ we get ($L\sim \xi_0$)

\begin{equation}
\delta I_1=\epsilon
\frac{e\Delta}{\hbar}\frac{L}{\xi_0}\frac{\cos^2(\phi/2)}{\sqrt{1-D\sin^2(\phi/2)}}.
\end{equation}

In the long junction limit ($L \gg \xi_0$) the current is given by
putting Eq. (\ref{longdedfi}) into (\ref{iinj}),

\begin{equation}
I_1=\epsilon\frac{e}{\hbar}\frac{\hbar
v_F}{2L}\frac{\cos^2(\phi/2)}{1-D\sin^2(\phi/2)}\sum_{n=0}^{N}[h(E_n^-)+h(E_n^+)
].
\label{iinjlong}
\end{equation}
The current step at zero temperature is

\begin{equation}
\delta I_1=\epsilon\frac{e}{\hbar}\frac{\hbar
v_F}{L}\frac{\cos^2(\phi/2)}{1-D\sin^2(\phi/2)}.
\end{equation}
For high temperatures $kT \gg \hbar v_F/L$, the sum (\ref{iinjlong}) can be
converted to an integral, just as for the anomalous current
(\ref{sumint}), which gives the current for $eV<\Delta$

\begin{equation}
I_1=\epsilon\frac{4e}{h}\frac{\cos^2(\phi/2)}{1-D\sin^2(\phi/2)} f(V,T).
\label{iinjhight}
\end{equation}
where $f(V,T)$ is given by Eq. (\ref{fdef}). The $IVC$ thus becomes
linear for $eV<\Delta$, with the slope independent of length and
temperature, as is seen in Fig. \ref{injection}. All information about
individual Andreev levels is washed out. 

The effect of asymmetry on the injection current is drastic in the
limit of a long junction with large asymmetry $L\gg l\gg \xi_0$, just
as for the asymmetric anomalous current. This shows the strong
relationship between the two currents. The injection current is given
by inserting Eq. (\ref{injcurrdens}) into (\ref{i1def})

\begin{equation}
I_1=\epsilon\frac{e}{\hbar}\frac{\hbar v_F}{L}\sum_{n=0}^{N}\frac{D
\sin^2(\phi)+R\sin^2\chi}{1-(D\cos\phi+R\cos\chi)^2}(n^e-n^h).
\end{equation}

Averaging over periods and summing up the filling factors, just as in
the case of the anomalous current [see Eq. (\ref{ianassum})], the
injection current becomes

\begin{eqnarray}
&&I_1=\epsilon\frac{e}{\hbar}\frac{8}{R\pi}  \left[1-\sqrt{D} \right.
\nonumber \\ && \left. \times
\left(\frac{|\sin(\phi/2)|^3}{\sqrt{1-D\cos^2(\phi/2)}}+\frac{|\cos(\phi/2)|^3}{\sqrt{1-D\sin^2(\phi/2)}}\right)\right] f(V,T). \nonumber \\
\label{iinjasym}
\end{eqnarray}
The injection current in this limit does not depend on either the
length $L$ or the asymmetry $l$. It is $\pi$-periodic,
$I_1(\phi+\pi)=I_1(\phi)$, for the same reasons as discussed above for
the asymmetric anomalous current. The implications of the $\pi$
periodicity for the conductance is discussed below.

\subsection{Conductance}

For the conductance, we discuss the whole range of coupling parameters
$\epsilon$, not only the weak coupling limit. The conductance is defined

\begin{eqnarray}
&&G(V,\phi)= \frac{dI_1}{dV}
\frac{d}{dV}(n^e-n^h)
 \nonumber \\
&&= \frac{e}{kT} \int_{-\infty}^{\infty}\frac{i_1^-}{2}
[\cosh^{-2}\frac{(E+eV)}{2kT}+ \cosh^{-2}\frac{(E-eV)}{2kT}].
\end{eqnarray}
At zero temperature $T=0$ the conductance for a symmetric
junction $l=0$ can be written \cite{Nakano} for $eV<\Delta$,
noting that $i_1^-(E)=i_1^-(-E)$,

\begin{equation}
G(V,\phi)=\epsilon^2 \frac{4e^2}{h}\frac{4
\cos^2(\phi/2)\sin^2\theta}{[(1-\epsilon)\cos2\theta-R-D\cos\phi]^2+
\epsilon^2\sin^2 2\theta}.
\label{conduct}
\end{equation}
At a voltage fulfilling the relation $1-\cos 2
\theta=D(1-\cos\phi)/\sqrt{1-2\epsilon}$, which is exactly at an
Andreev resonance, and a phase difference $\phi=0$ mod $2\pi$, the
conductance is $G=4e^2/h$, which is the conductance for a perfect
NS-interface.  This holds for any transparency $D$, length $L$ and
coupling constant $\epsilon$.

For $L=0$ we get $\cos\theta = eV/\Delta$ to be inserted into
(\ref{conduct}). This gives rise to a peak in the conductance at the
voltage $eV\approx E_0^-$, the energy of the single Andreev
resonance. In the long limit $(L\gg \xi_0)$ for $E \ll \Delta$ we get
$\theta=\pi/2-eV/E$ which results in conductance oscillations as a
function of applied voltage
\cite{Morpurgo97b,Volkzait,Dimoulas,Lesovik}, with a period $\pi \hbar
v_F/L$, the distance between the pairs of Andreev levels. This is made
clear by taking the derivative of current with respect to voltage in
Fig. \ref{injection}.

\begin{figure}
\centerline{\psfig{figure=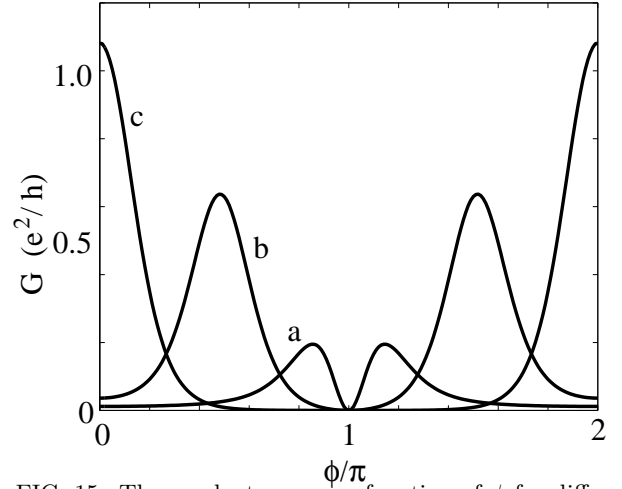,height=6.5cm}}
\caption{The conductance as a function of $\phi$ for different voltages
(a) $eV=0$, (b) $eV=0.075\Delta$ and c) $eV=0.15\Delta$ at temperature
$kT=0.01\Delta$. $D=0.9$, $\epsilon=0.05$ and $L=20\xi_0$.}
\label{condflucv}
\end{figure}

The conductance vs phase relation $G(V,\phi)$ is also altered when
voltage is applied.  For voltages around $eV/(\hbar v_F/L) \approx n
\pi$, i.e at an energy between the pairs of Andreev resonances, the
conductance has a maximum around $\phi\approx\pi$ and a local minimum
at $\phi=0$ [apart from the absolute minimum at $\phi=\pi$ due to the
factor $\cos^2(\phi/2)$ in Eq. (\ref{conduct})]. For $eV/(\hbar v_F/L)
\approx\pi/2+n\pi$, i.e at an energy between the two Andreev
resonances in the pair, the maximum shifts to $\phi=0$ and the minimum
to $\phi=\pi$ [see Fig. (\ref{condflucv})]. This behavior has
recently been observed in both ballistic \cite{Morpurgo97b} and
quasiballistic \cite{Dimoulas} junctions. It has also been predicted
for diffusive junctions. \cite{Volkzait,Leadbeater} This voltage
dependence of the conductance holds for $kT \ll \hbar v_F/L$.

For the additional condition zero voltage $V=0$, the expression
(\ref{conduct}) reduces to

\begin{equation}
G(0,\phi)= \epsilon^2 \frac{2e^2}{h}\frac{2\cos^2(\phi/2)}{R+D\cos^2(\phi/2)}.
\end{equation}
For $D\leq R$ the maximum conductance has a universal magnitude
$G_{max}(0)=G_N\epsilon/(1-\epsilon)^2$ where $G_N=\epsilon 4e^2/h$ is
the conductance of the junction in the normal state.

\begin{figure}
\centerline{\psfig{figure=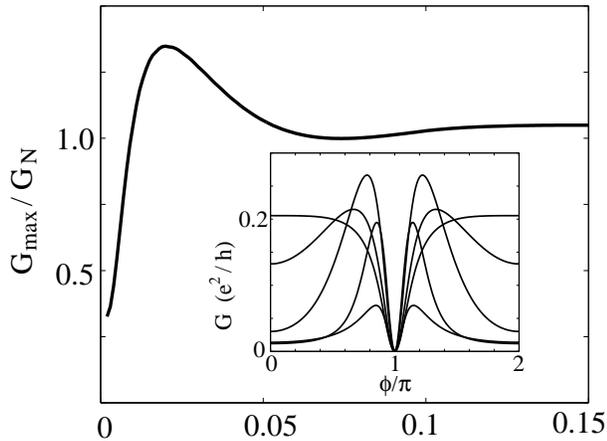,height=6.5cm}}
\caption{The maximum conductance $G_{max}/G_N$ as a function of $T$
for zero voltage. $L=10\xi_0$, $D=0.9$ and $\epsilon=0.05$. At $T=0$
the maximum conductance is $G_{max} \ll G_N$. It increases with
increasing $T$, reaches a maximum around $kT\sim \hbar v_F/L$, drops
again for $kT>\hbar v_F/L$ but saturates at a constant value
$G_{max}=G_N$ for $kT \gg \hbar v_F/L$. Inset: The conductance as a function
of $\phi$ for different temperatures $kT=0, 0.01\Delta, 0.025\Delta,
0.05\Delta$ and $0.1\Delta$ at zero voltage $V=0$. Increasing
temperature from bottom to top at $\phi=0$.  $D=0.9$, $\epsilon=0.05$
and $L=20\xi_0$.}
\label{condtemp}
\end{figure}

When the coupling is weak, $\epsilon \ll 1$, the maximum conductance
is $G_{max} \sim \epsilon G_N$, i.e much smaller than the normal
conductance. In this limit the Andreev resonances are sharp and there
are no available Andreev states at $E_F$, because the scattering at
the three lead connection point opens up a gap in the Andreev spectrum
at $\phi=\pi$ (see Fig. \ref{asandr}). The conductance is thus
strongly suppressed.  The conductance rapidly increases with voltage
and temperature and has a maximum at $eV,kT\sim \hbar v_F/L$. This
happens because electrons (holes) with energies $E>0$ ($E<0$) tunnel
into the first resonant Andreev state at finite energy. This gives
rise to a finite energy conductance peak \cite{Kastalsky} (see
Fig. \ref{condtemp}). The maximum conductance $G_{max}$ is plotted as
a function of temperature in Fig.  \ref{condtemp} (note that the
minimum conductance always is zero).

The conductance as a function of phase difference $\phi$ behaves
differently for different temperatures $T$, as is shown in the inset
in Fig. \ref{condtemp}. For $kT\gg \hbar v_F/L$, a limit only accessible for
the long junction $L\gg\xi_0$, the maxima of the conductance around
$\phi\approx\pi$ are shifted to a maximum at $\phi=0$ mod $2\pi$. This
holds independent of voltage applied. A similar effect in a
quasiballistic system has been reported by Dimoulas et al.\cite{Dimoulas}

In the weak coupling limit $\epsilon \ll 1$ we can use expression
(\ref{iinjhight}) to get the conductance in the long limit for $\hbar
v_F/L \ll kT\ll \Delta$ and $eV<\Delta$

\begin{equation}
G(\phi)=\epsilon \frac{4e^2}{h}\frac{\cos^2(\phi/2)}{1-D\sin^2(\phi/2)}
\label{condsimp}
\end{equation}

which is independent of voltage, temperature and length of the
junction and has a maximum $G_{max}=G_N$ at $\phi=0$ mod $2\pi$.

In the same limit we get the conductance in the asymmetric junction
from expression (\ref{iinjasym})

\begin{eqnarray}
&&G(\phi)=\epsilon\frac{e}{\hbar}\frac{8}{R\pi} \left[1-\sqrt{D}
\right.  \nonumber \\
&& \left. \times
\left(\frac{|\sin(\phi/2)|^3}{\sqrt{1-D\cos^2(\phi/2)}}+\frac{|\cos(\phi/2)|^3}{\sqrt{1-D\sin^2(\phi/2)}}\right)\right].
\label{condas}
\end{eqnarray}

It is $\pi$-periodic and has a maximum for $\phi=\pi(n+1/2)$ and a
minimum for $\phi=\pi n$. This $\pi$-periodicity can be qualitatively
explained by considering the lowest order paths giving rise to the
conductance.

\begin{figure}
\centerline{\psfig{figure=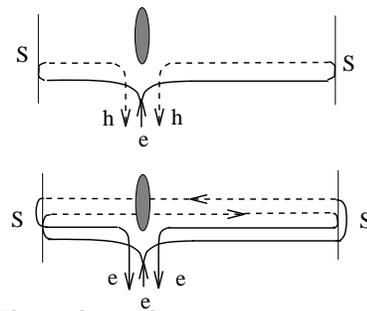,height=4cm}}
\caption{The paths in the asymmetric junction giving the first order terms
of the conductance. Electrons are drawn with solid lines, holes with
dashed. The grey ellipse denotes the effective scatterer due to the three
lead connection. The upper paths give rise to a $2\pi$ periodic component
of the current, suppressed at finite temperature $kT\gg \hbar v_F/L$. The lower
paths, time reversed, give rise to a $\pi$ periodic component of the
current, not suppressed by temperature.}
\label{asym}
\end{figure}

The upper paths in Fig. \ref{asym}, corresponding to an injected
electron and giving rise to an outgoing hole, produce a current of the
order $i_1\sim
|e^{i(\phi_L+\beta_2)}+e^{i(\phi_R+\beta_3)}|^2=2+2\cos(\phi+\chi)$. This
part of the current is $2\pi$ periodic in the phase, rapidly
oscillating in energy with a period $\hbar v_F/l$. It is washed out
when summing up the levels in a long junction at finite
temperature. The lower paths in Fig.  \ref{asym}, corresponding to an
injected electron giving rise to an outgoing electron, produce a
current of the order $i_1 \sim
|d^*e^{i(\phi+\beta)}+d^*e^{i(-\phi+\beta)}|^2=D(2+2\cos2\phi)$. This
part of the current is $\pi$ periodic in phase and not sensitive to
asymmetry.

The discussion about the periodicity of the conductance oscillations
with respect to phase goes back to the early eighties. A
$\pi$-periodic contribution to the weak localization correction to the
conductance in a similar system was predicted by Spivak et
al.\cite{Spivak} and discussed further by Altshuler et
al. \cite{Altshuler} It has been shown in numerical simulations for a
structure similar to ours that the full conductance, i.e not only the
weak localization contribution, might become $\pi$-periodic at finite
temperatures. \cite{Hui} A large $\pi$-periodic conductance
oscillation with phase was also observed in diffusive samples
\cite{Petrashov}. Whether the explanation to the crossover from $2\pi$
to $\pi$ periodicity with increased temperature discussed above can
account for these observations remains to be investigated.

\section{Four terminal junction}

One problem with the three terminal junction is, as discussed above,
that it is not possible to separate the injection current from the
Josephson current in a clear way for arbitrary coupling $\epsilon$. In
a four terminal
junction\cite{Bagwell,Volkov95,Kadigrobov2,Petrashov4,Nazstoof} this
is possible under certain conditions, which makes it interesting to
discuss this configuration separately.

\begin{figure}
\centerline{\psfig{figure=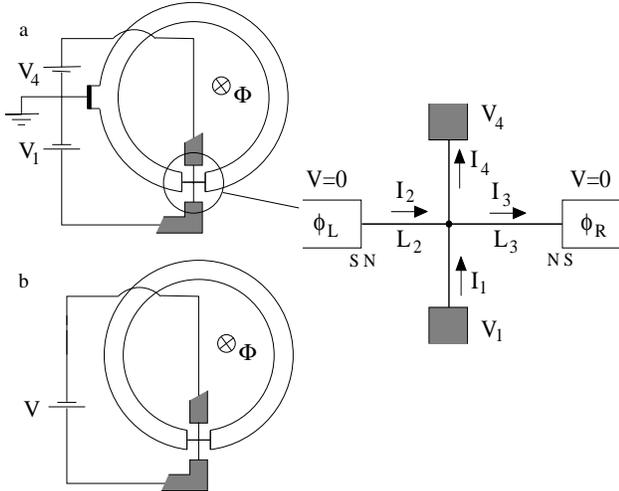,height=6.5cm}}
\caption{Two different setups of the four terminal junction. In (a) the
normal reservoirs are biased independently at $V_1$ and $V_4$ with
respect to the superconducting loop (grounded), in (b) only the potential
difference between the normal reservoirs, $V$, is determined. In the
right figure, a close up of the junction area is shown, with the
direction of the currents showed with arrows}
\label{fourterm}
\end{figure}

We consider two different types of junction configurations (see
Fig. \ref{fourterm}). The upper junction is a straightforward
extension of the three terminal device pictured in
Fig. \ref{juncfig}. Two normal reservoirs are connected to the normal part of the junction. The reservoirs are then connected to
the grounded superconducting loop via voltage sources biased at
$V_1$ and $V_4$ respectively.

The current injected from a normal reservoir is split at the
connection point. One part of the current flows through the junction
directly into the other reservoir and the other part of the current is
divided between the leads $2$ and $3$. In the general case, the
currents in lead $1$ and $4$ are not equal $I_1 \neq I_4$. However, by
adjusting the potentials $V_1$ and $V_4$, the currents in the vertical
lead can be put equal and from current conservation at the connection
point it follows $I_2=I_3$ and we have a clear separation between the
injection current $I_1=I_4$ and the Josephson current circulating in
the superconducting loop $I_2=I_3$.

In the lower junction, this separation follows directly from current
conservation at the connection point, since $I_2=I_3$. In this
junction a bias $V$ is applied between the normal reservoirs, which
are only connected to the superconducting loop via the four lead
connection point. We define the potential of the superconducting loop
in this junction to be zero and the potentials of the normal
reservoirs to $V_1$ and $V_4$ resepectively, just as for the upper
junction. Our biasing arrangement then gives $V=V_1-V_4$. The
condition of current conservation, $I_1(V_1,V_4)=I_4(V_1,V_4)$, gives
a second condition on $V_1$ and $V_4$. With these definitions we can
calculate the current in both junctions in the same way.

The cross-shaped connection point is modelled by the scattering matrix

\begin{equation}
S=\left( \begin{array}{llll}
         r_{\perp} & \sqrt\epsilon & \sqrt\epsilon & d_{\perp}\\
         \sqrt\epsilon & r & d & \sqrt\epsilon\\
         \sqrt\epsilon & d & r & \sqrt\epsilon \\
         d_{\perp} & \sqrt\epsilon & \sqrt\epsilon & r_{\perp}\\
\end{array} \right)
\end{equation}
where the $\epsilon$ describes the coupling of the SNS junction to the
vertical normal lead ($0\leq \epsilon\leq 0.25$). The horisontal scattering
amplitudes now obey the relations $Re(rd^*)=-\epsilon$ and
$D+R=1-2\epsilon$. The same holds for the vertical scattering
amplitudes $r_{\perp}$ and $d_{\perp}$.

The current densities $i_{j}^{e(h),1(4)}$, with the upper index $1$ or
$4$ denoting the lead from which the quasiparticles are injected, are
calculated in the same way as in the case of the three terminal
junction. Due to the symmetry of the scattering matrix, quasiparticles
injected from leads $1$ or $4$ give rise to the same current density
in leads $2$ and $3$, i.e $i_{2(3)}^{e(h),1}=i_{2(3)}^{e(h),4}$.

The expressions for the sum and anomalous current densities become
very similar to the three terminal expressions [see
Eq. (\ref{sumcurr}) and (\ref{avdiffcurr})], i.e one just changes
$\epsilon \rightarrow 2\epsilon$ [also changing $Z=Z(\epsilon
\rightarrow 2\epsilon)$] and divides by two, noting that
$i^{+,1}=i^{+,4}$ and $i_a^{1}=i_a^{4}$. Neither the vertical
transparency $D_{\perp}$ nor the reflectivity $R_{\perp}$ thus appear
explicitly in these expressions. The factor one half simply reflects
that there are {\it two} normal leads connected to the normal part of
the junction. In the limit of weak coupling $\epsilon \ll 1$, the sum
of the current densities from both normal reservoirs is equal to the
current density from the single normal reservoir in the three terminal
junction, $i^{+,1}+i^{+,4}=i^+$ and $i_a^1+i_a=i_a$ (simply reflecting
that one cannot create more Josephson current by adding more normal
leads).

In this weak coupling limit the current in the horisontal lead
$I=I_2=I_3$ is given by

\begin{equation}
I=I_{eq}+\frac{1}{2}[I_r(V_1)+I_r(V_4)]+\frac{1}{2}[I_a(V_1)+I_a(V_4)]
\end{equation}
with $I_{eq},I_r$ and $I_a$ the same as in the three terminal case,
given by Eq. (\ref{irbound})-(\ref{sumboundstate}). Noting the
relations $I_r(-V)=I_r(V)$ and $I_a(-V)=-I_a(V)$, we see that (i) for
bias $V_1+V_4=0$ the anomalous current is zero, and (ii) for
$V_1-V_4=0$ the regular current is zero. We can thus control the
regular and anomalous currents in the upper junction in
Fig. \ref{fourterm} independently by adjusting either the potential
difference $V_1-V_4$ or the sum $V_1+V_4$ between the normal
reservoirs, keeping the other quantity constant.

\subsection{Injection current and conductance}

The injection currents $I_1$ and $I_4$ in the four terminal device is
qualitatively different from the injection current in the three
terminal device, since in the four terminal junction the injected
quasiparticles from one normal reservoir can travel directly through
the junction to the other normal reservoir.

The current in leads $j=1,4$ can be written [see Eq. (\ref{totcurr})]

\begin{eqnarray}
I_j&=&\int_{-\infty}^{\infty}\ dE \left[\frac{i_j^{+,1}}{2}(n^{e,1}+n^{h,1})+
\frac{i_j^{-,1}}{2}(n^{e,1}-n^{h,1})+ \right. \nonumber \\
&& \left.
+\frac{i_j^{+,4}}{2}(n^{e,4}+n^{h,4})+\frac{i_j^{-,4}}{2}(n^{e,4}-n^{h,4})
\right]
\label{fourinj}
\end{eqnarray}

The symmetry of the junction gives that $i^{-,1}_4=-i^{-,4}_1$ and
$i^{-,4}_4=-i^{-,1}_1$. This leads to that for $V_1=-V_4=V/2$ the
currents in the vertical leads are equal, i.e $I_{inj}=I_1=I_4$ and
thus no injection current flows into the superconductors. In this case
Eq. (\ref{fourinj}) reduces to

\begin{equation}
I_{inj}=\int_{-\infty}^{\infty} dE \left[ \frac{i_1^{-,1}-i_1^{-,4}}{2}(n^{e,1}-n^{h,1})\right].
\end{equation}

Considering the simplest case with a symmetric junction $l=0$ without
barriers at the NS-interfaces. The injection current density is given
by straightforward calculation

\begin{eqnarray}
&&i_1^{-,1}-i_1^{-,4}=4\frac{e}{h}\left\{D_{\perp}+\epsilon+\frac{\epsilon}{Z_f
}\left[\left([1-2\epsilon]\cos2\theta-D\cos\phi \right. \right. \right.
\nonumber \\
&&\left. \left. \left. -R
\right)[R_{\perp}-D_{\perp}-2\mbox{Re}[d(d_{\perp}-r_{\perp})]\sin^2(\phi/2)]+\left(\sin^22\theta\nonumber \right. \right. \right. \\
&&\left. \left. \left. -1+2\epsilon+(D\cos\phi+R)\cos2\theta
\right)(R_{\perp}-D_{\perp})\right]\right\}.
\end{eqnarray}
Some general comments can be made about the conductance
$G=dI_{inj}/dV$. When the vertical and horisontal leads are decoupled
($\epsilon \rightarrow 0$), the conductance reduces to
$G=(e^2/h)D_{\perp}$, the conductance of the normal vertical
channel. For finite coupling, an additional term is added to the
conductance $\delta G\sim\epsilon$. This additional term $\delta G$ is
dependent on the phase difference $\phi$, but it is also, unlike for
the three terminal junction, dependent on the scattering amplitudes
$r,d,r_{\perp}$ and $d_{\perp}$, i.e not only the scattering
probabilities $R,D,R_{\perp}$ and $D_{\perp}$. This becomes clear when
we note that we can rewrite the expression
$\mbox{Re}[d(d_{\perp}-r_{\perp})]=1/2[R_{\perp}-D_{\perp}+([R_{\perp}-D_{\perp}][R-D]-4\mbox{Im}[r_{\perp}d_{\perp}^*]\mbox{Im}[rd^*])/(1-4\epsilon)]$,
i.e dependent on $\sigma \sigma_{\perp}$, just like the anomalous
current. This contribution can be explained qualitatively by
interference between quasiparticle paths where one path describes
scattering in the vertical direction and the other one in the
horisontal direction, thereafter leaving the junction.

\begin{figure}
\centerline{\psfig{figure=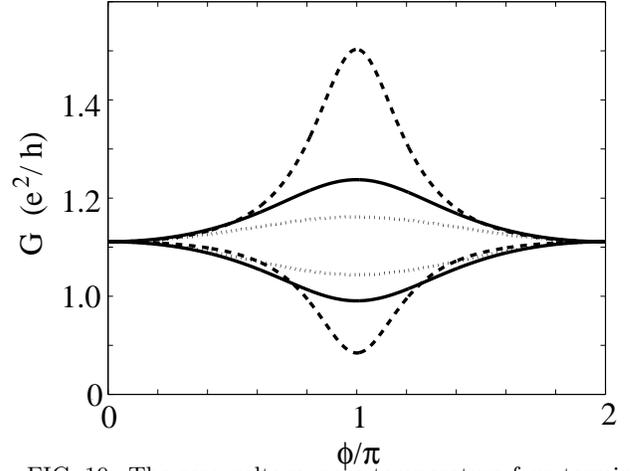,height=6.5cm}}
\caption{The zero voltage, zero temperature four terminal conductance
$G$ as a function of phase difference $\phi$ for different horisontal
transparency $D=0.2$ (dashed), $D=0.5$ (solid) and $D=0.8$
(dotted). The product $\sigma \sigma_{\perp}=+1$ for increase of
conductance due to finite phase difference, $\sigma \sigma_{\perp}=-1$
for decrease. The vertical transparency is $D_{\perp}=0.5$
and $\epsilon=0.05$.}
\label{fourtermcond}
\end{figure}

In the case of zero temperature and voltage, the conductance becomes

\begin{eqnarray}
G&=&\frac{2e^2}{h}\left[D_{\perp}\right. \nonumber \\ && \left.
+\epsilon\left(1+\frac{D_{\perp}-R_{\perp}+\sin^2(\phi/2)\mbox{Re}[d(d_{\perp}-r
_{\perp})]}{R+D\cos^2(\phi/2)}\right)\right]
\end{eqnarray}
which is independent of the length $L$ of the junction. The
conductance at zero phase difference given by
$G(\phi=0)=(2e^2/h)[D_{\perp}/(1-2\epsilon)]$. From this
value, the conductance then increases or decreases, depending on the
phases of the scattering amplitudes, monotonically with
$\phi\rightarrow\pi$, as is seen in Fig. \ref{fourtermcond}.

For finite energies, the injected quasiparticles tunnel into the
Andreev resonances and there is a finite bias anomaly of the
conductance. However, there is not always a peak in the
conductance. This is made clear by considering the conductance at
finite voltage, zero temperature and zero phase difference (to avoid
dependence on $\sigma \sigma_{\perp}$), given by 

\begin{equation}
G=\frac{2e^2}{h} \left(D_{\perp}+\epsilon\left[1+\frac{(R_{\perp}-D_{\perp})[\cos^2\theta-(1-2\epsilon)]}{(1-2\epsilon)^2\sin^2 \theta+4\epsilon^2\cos^2\theta} \right]\right).
\end{equation}

In the weak coupling limit, $\epsilon \ll 1$, for
$R_{\perp}>D_{\perp}$, there is an increase in the conductance for
finite energies, just as for the three terminal device. For
$R_{\perp}<D_{\perp}$, i.e for highly transmissive junctions in the
vertical direction, however, there is a {\em decrease} in conductance
for finite energies. The presence of Andreev resonances thus decreases
the probability of a quasiparticle to be transmitted through the
junction.\cite{Morpurgo97b}

Other properties of the conductance are similar to the three terminal
junction, also taking the scatting phases into account (via $\sigma$
and $\sigma_{\perp}$). The conductance is periodic in voltage with a
period $\pi \hbar v_F/L$, the distance between the pairs of Andreev
resonances, for temperatures lower than this inter pair distance $kT\ll
\hbar v_F/L$. For such low temperatures the phase dependence of the
conductance is also dependent on the voltage. For high temperatures
$kT \gg \hbar v_F/L$, all the features of the individual Andreev
levels are washed out and the conductance becomes independent on
junction length and applied voltage $eV<\Delta$. For $eV>\Delta$, the
conductance is equal to the normal conductance of the junction
$G_N=(2e^2/h)(D_{\perp}+2\epsilon)$.

\section{Conclusions}

We have analyzed the equilibrium and nonequilibrium Josephson currents
and conductance in a ballistic, multiterminal, single mode SNS
junction. The nonequilibrium is created by means of quasiparticle
injection from a normal reservoir connected to the normal part of the
junction. By applying a voltage $V$ to the normal reservoir, up to the
superconducting gap $\Delta$, the equilibrium current of a short
junction $L\ll\xi_0$ can be suppressed. When the junction is longer
$L\geq \xi_0$, the direction of the Josephson current changes sign
as a function of applied voltage. For a junction longer then the
thermal length $L \gg \xi_T$, the {\em equilibrium} Josephson current
is exponentially small. The {\em nonequilibrium} Josephson current in
this regime is dominated by the {\em anomalous current}, arising from
the modification of the current carrying Andreev states due to
coupling to the normal reservoir. This anomalous current scales
linearly with applied voltage and saturates at a magnitude of the
order of the equilibrium current carried by a short junction, $I\sim
e\Delta/\hbar$.

The conductance oscillates as a function of the phase difference
$\phi$ between the superconductors, with a period of $2\pi$ in a
symmetric junction. The position of the conductance minima, $\phi=0$
or $\pi$, is dependent both on applied voltage and temperature. The
conductance exhibits a finite bias anomaly, at $eV \sim \hbar v_F/L$,
the position of the first current carrying Andreev level.

Asymmetric injection gives rise to oscillations of all currents on the
scale of $eV\sim \hbar v_F/l$ where $l$ is the length difference
between the two leads. At temperatures above this energy, these
oscillations are smeared and a we get renormalized anomalous and
injection currents that are $\pi$-periodic.

Introduction of barriers at the NS-interfaces give a strong length
dependence of all currents, governed by the Breit Wigner resonances between
the normal barriers. There are resonant current peaks at lengths where
the normal electron and hole resonances cross. 

Connecting a second normal reservoir to the normal part of the
junction allows a clear separation between the injection current,
flowing between the two normal reservoirs, and the Josephson current,
flowing between the superconductors. 

\acknowledgements

This work has been supported by research grants from NFR, TFR, NUTEK
(Sweden) and by a NEDO International Joint Research Grant (Japan).

\appendix

\section{Continuum state current}

Here we present formulas for the continuum current for a symmetric
($l=0$) three terminal junction without barriers at the NS-interfaces.
The continuum current consists of particles injected from both the
normal reservoir and the superconducting reservoirs.  The current
density in lead 2 from all injected quasiparticles from the
superconductors is

\begin{eqnarray}
i^s_2&=&\frac{e}{h}\frac{2\sin\beta\sinh\gamma_c}{Z_c}\left\{\sin\phi\left[(4D-2
D \epsilon-\epsilon^2)\cosh\gamma_c \right. \right. \nonumber \\
&& \left. \left. -2D\epsilon e^{-\gamma_c}\right]-4\sigma
\epsilon\sqrt{RD-\epsilon^2}\sin^2(\phi/2)\cosh\gamma_c\right\},
\end{eqnarray}
where $Z_c=[\cos\beta(\cosh 2\gamma_c (1-\epsilon)+\epsilon\sinh
2\gamma_c)-R-D\cos\phi]^2+[\sin\beta(\sinh 2\gamma_c (1-\epsilon)+\epsilon
\cosh 2\gamma_c)]^2$ and $\gamma_c=\mbox{arccosh}(E/\Delta)$. In lead 3 we
get $i^s_3(\phi)=-i^s_2(-\phi)$.
This current density is an oscillating function of energy with
largest amplitude for energies close to $E=\Delta$ and is given at negative
energies by $i^s_j(E)=-i^s_j(-E)$.

For the particles injected from the normal reservoir, the sum
current in lead 2 becomes

\begin{eqnarray}
i^+_2&=&2\frac{e}{h}\frac{\epsilon
\sin\beta\cosh\gamma_c}{Z_c}\left\{\sin\phi
\left[2D\cosh\gamma_c+\epsilon\sinh\gamma_c \right] \right. \nonumber \\
&& \left. +4\sigma \sqrt{RD-\epsilon^2/4}\sin^2(\phi/2)\sinh\gamma_c\right\}
\end{eqnarray}
with $i_3^+(\phi)=-i_2^+(-\phi)$ in lead 3. The difference current in lead
2 has the form

\begin{eqnarray}
i^-_2&=&2\frac{e}{h}\frac{\epsilon
\cosh\gamma_c}{Z_c}\left\{-\cosh\gamma_c[\epsilon\cos\phi+2\sigma\sqrt{RD-\epsilon^2/4}\sin\phi]
\right.  \nonumber \\ &&
\left. +(1-\epsilon)[\sinh\gamma_c-\sinh3\gamma_c]-\epsilon\cosh3\gamma_c
\right. \nonumber \\ &&
\left. +\cos\beta [2\sinh\gamma_c(R+D\cos\phi) \right.
\nonumber \\ &&
\left.
+\cosh\gamma_c(\epsilon(1+\cos\phi)+2\sigma\sqrt{RD-\epsilon^2/4}\sin\phi)]\right\}
\label{contdiffcurr}
\end{eqnarray}
with $i_3^-(\phi)=-i_2^-(-\phi)$ in lead 3. For negative energies we
get $i^+_j(E)=-i^+_j(-E)$ and $i^-_j(E)=i^-_j(-E)$.

\section{Spectral densities of currents in the weak coupling limit}

In this appendix we analyze the central quantity in the current
density expressions (\ref{sumcurr})-(\ref{injdiffcurr}), given by

\begin{equation}
\frac{\epsilon}{Z}=\frac{\epsilon}{[(1-\epsilon)\cos 2\theta-R\cos \chi-D\cos\phi]^2+\epsilon^2
\sin^2 2\theta},
\end{equation}
in the limit of zero coupling $\epsilon \rightarrow 0$. We can conveniently rewrite
\begin{equation}
\frac{\epsilon}{Z}=\frac{1}{\sin^2 2\theta}\frac{\epsilon}{F^2+\epsilon^2}
\label{epsgotozero}
\end{equation}
with $F(E,\phi)=[(1-\epsilon)\cos 2\theta-R\cos \chi-D\cos\phi]/\sin 2\theta$. In the limit of zero coupling the expression becomes

\begin{equation}
\lim_{\epsilon\rightarrow 0}
\frac{\epsilon}{F^2+\epsilon^2}=\pi \delta(F)=\sum_{n,\pm} \pi \left| \frac{\partial}{\partial E} F \right|^{-1} \delta(E-E_n^{\pm})
\label{lastref}
\end{equation}
with the energies $E_n^{\pm}$, given by 

\begin{equation}
\cos 2\theta-R\cos\chi-D\cos\phi = 0,
\label{disprel}
\end{equation}
being the energies of the bound Andreev states. By rewriting 

\begin{equation}
\frac{\partial}{\partial E} F = \frac{d\phi}{dE}\frac{\partial}{\partial \phi} F=\frac{d\phi}{dE} \frac{D \sin \phi}{\sin 2\theta}
\end{equation}
the expression (\ref{lastref}) becomes 

\begin{equation}
\lim_{\epsilon\rightarrow 0} \frac{\epsilon}{Z}=\sum_{n,\pm} \frac{\pi}{D|\sin \phi \sin 2\theta|}\left|\frac{dE}{d\phi}\right| \delta(E-E_n^{\pm}).
\label{epszeta}
\end{equation}
From Eq. (\ref{disprel}), the derivative of energy with respect to phase becomes

\begin{equation}
\frac{dE}{d\phi} = -\frac{D\sin\phi}{2\sin2\theta
\left(
\frac{1}{\sqrt{\Delta^2-E^2}} + \frac{L}{\hbar v_F} +
\frac{l}{\hbar v_F}R\frac{\sin\chi}{\sin2\theta}
\right)}.
\label{dedfi}
\end{equation}
Using the relation (\ref{disprel}) and the fact that $R+D=1$ we can rewrite 
\begin{eqnarray}
&&|\sin 2\theta|=\sqrt{(D+R)^2-(D \cos \phi +R\cos \chi)^2} \nonumber \\
&& = \sqrt{D^2 \sin^2\phi+R^2\sin^2 \chi +2DR(1-\cos\phi\cos\chi)}.
\end{eqnarray}
which shows that $|\sin 2\theta| > R |\sin \chi|$. This gives that, since $L \geq l$ by definition, the factor in the parenthesis in the denominator in Eq. (\ref{dedfi}) is always positive. The relation  

\begin{equation}
\mbox{sgn}\left(\frac{dE}{d\phi} \sin\phi \sin2\theta\right) = -1
\label{signrel}
\end{equation}
then follows from Eq. (\ref{dedfi}).
The Eqs. (\ref{epszeta}) and (\ref{signrel}) are the technical result of this appendix. 

\section{Currents for different lengths}

Here we list all expressions for the partial currents in different length
limits. The junction considered is a symmetric ($l=0$) three terminal junction
without barriers at the NS-interfaces, in the weak coupling limit $\epsilon
\ll 1$.

\subsection{Short limit $(L=0)$}

\begin{equation}
I_{eq}^{b}=\frac{e\Delta}{\hbar}\frac{D\sin\phi}{2\sqrt{1-D\sin^2(\phi/2)}}\tanh
(E_0^-/2kT),
\end{equation}

\begin{equation}
I_r=\frac{e\Delta}{\hbar}\frac{D\sin\phi}{4\sqrt{1-D\sin^2(\phi/2)}}g(E_0),
\end{equation}

\begin{equation}
I_a=\frac{e}{\hbar}\frac{\sigma D\sqrt{R}\sin\phi
|\sin(\phi/2)|}{1-D\sin^2(\phi/2)}h(E_0),
\end{equation}

\begin{equation}
I_1=\frac{e}{\hbar}\frac{\epsilon
\sqrt{D}|\sin(\phi/2)|}{1-D\sin^2(\phi/2)}h(E_0).
\end{equation}

When there are two Andreev levels, with $1\ll D$, $\beta/2>\sqrt{D}$ and
$E_0^+\approx E_0^-\approx \Delta$, the currents become:

\begin{eqnarray}
I_{eq}^{b}&=&\frac{e\Delta}{\hbar}\frac{L}{\xi_0}\frac{\sqrt{D}\sin(\phi)}{2|\sin(\phi/2)|\sqrt{1-D\sin^2(\phi/2)}} \nonumber \\
&& \left[\tanh(E_0^-/2kT)-\tanh(E_0^+/2kT)\right],
\end{eqnarray}

\begin{equation}
I_r=\frac{e\Delta}{\hbar}\frac{L}{\xi_0}\frac{\sqrt{D}\sin(\phi)}{4|\sin(\phi/2)
|\sqrt{1-D\sin^2(\phi/2)}}[g(E_0^+)-g(E_0^-)],
\end{equation}

\begin{equation}
I_a=\sigma
\frac{e\Delta}{\hbar}\frac{L}{2\xi_0}\frac{\sqrt{RD}\sin\phi}{1-D\sin^2(\phi/2)}
[h(E_0^+)+h(E_0^-)],
\end{equation}

\begin{equation}
I_1=\epsilon
\frac{e\Delta}{\hbar}\frac{L}{\xi_0}\frac{\cos^2(\phi/2)}{\sqrt{1-D\sin^2(\phi/2
)}}[h(E_0^-)+h(E_0^+)].
\end{equation}

\subsection{Long limit $(L \ll \xi_0)$}

\begin{eqnarray}
I_{eq}^{b}&=&\frac{e}{\hbar}\frac{\hbar
v_F}{L}\frac{\sqrt{D}\sin(\phi)}{2|\sin(\phi/2)|\sqrt{1-D\sin^2(\phi/2)}} \
\nonumber \\
&& \times \left(\sum_{n=0}^{N}
\left[\tanh(E_n^-/2kT)-\tanh(E_n^+/2kT)\right]\right) \nonumber \\
&& +i^{*}\tanh(\Delta/2kT),
\end{eqnarray}

\begin{eqnarray}
I_r&=&\frac{e}{\hbar}\frac{\hbar
v_F}{2L}\frac{\sqrt{D}\cos(\phi/2)}{\sqrt{1-D\sin^2(\phi/2)}}\left(\sum_{n=0}^{N
}[g(E_n^-)-g(E_n^+)]\right) \ \nonumber \\
&& +\frac{i^*}{2}g(\Delta),
\end{eqnarray}

\begin{equation}
I_a=\frac{e}{\hbar}\frac{\hbar
v_F}{L}\frac{\sigma \sqrt{DR}\sin\phi}{2[1-D\sin^2(\phi/2)]}\sum_{n=0}^{N}[h(E_n^
+)+h(E_n^-)],
\end{equation}

\begin{equation}
I_1=\frac{e}{\hbar}\frac{\epsilon \hbar
v_F}{2L}\frac{\cos^2(\phi/2)}{1-D\sin^2(\phi/2)}\sum_{n=0}^{N}[h(E_n^-)+h(E_n^+)
].
\end{equation}

where $h(E)=\tanh[(E-eV)/2kT]-\tanh[(E+eV)/2kT]$ and
$g(E)=\tanh[(E+eV)/2kT]+\tanh[(E-eV)/2kT]-2\tanh(E/2kT)$.

\end{document}